\documentclass{article}

\usepackage{times,amsmath,amsfonts,amssymb,latexsym,textcomp}
\usepackage{color}
\usepackage{graphicx}
\usepackage{multirow}
\usepackage{bbm}
\usepackage{tabularx}
\usepackage{amsthm}
\usepackage{dsfont}
\usepackage{float}
\usepackage{array}
\usepackage{placeins}
\usepackage{MnSymbol}
\usepackage[electronic]{ifsym}
\usepackage[lofdepth,lotdepth]{subfig}

\usepackage[T1]{fontenc}
\usepackage[latin9]{inputenc}
\usepackage[english]{babel}

\begin{document}

\title{Contextuality, Pigeonholes, Cheshire Cats, Mean Kings, and Weak Values}
\author{Mordecai Waegell \footnote{caiwaegell@gmail.com},  Jeff Tollaksen \footnote{tollakse@chapman.edu} \\ \it{ Institute for Quantum Studies, Chapman University} \\ {\it Orange, CA, USA}}

\maketitle

\textbf{Abstract:} The structural connections between the Kochen-Specker (KS) theorem, pre- and post-selection (PPS) paradoxes, and anomalous weak values are explored in detail.  All PPS paradoxes, such as the 3-box paradox, the Quantum Cheshire Cat, and the Quantum Pigeonhole principle, construct a particular type of ontological model that assigns an eigenvalue to each observable (independent of context) of a system such that these assignments are consistent with the PPS.  It is shown that such an ontological model must be explicitly contextual in the sense of the KS theorem, or otherwise implies either a restriction on free random choice or explicitly retrocausal behavior.  We call such models PPS-contextual.  The structure of each paradox is always such that there are particular contexts of mutually commuting observables that violate the product rule or sum rule, when the ontological model is extended to include observables that are not measured during the experiment.  These paradoxes are counterfactual, in the sense that they are not directly observed, and also because the product and sum rules are always obeyed by projective measurements in actual experiments.  It is shown that by adopting an alternate ontological model, where all hidden variables are weak values (which are not always eigenvalues, but obey the sum rule by definition), the same contexts that presented the original paradox must also contain observables with anomalous weak values.  These anomalous weak values are not counterfactual because they can be probed through weak measurements on an ensemble of identically pre- and post-selected states, allowing this localized signature of KS contextuality to be experimentally observed.  The weak values of all observables of a system can in principle be measured during an experiment, making this model a promising candidate for describing PPS-contextual ontological `elements of reality.'  As a related issue, we show using the mathematical properties of weak values, that any KS set can be used to ensure that the Mean King always wins his game against the stranded physicist.


\subsection{Introduction}
 \subsubsection{Overview}
The main purpose of this paper is to point out a fundamental connection between Kochen-Specker contextuality \cite{KS}, the recently discussed quantum pigeonhole effect \cite{aharonov2014quantum, yu2014quantum}, and anomalous weak values \cite{aharonov1988result, aharonov1991complete, dressel2014weak, tollaksen2007pre, pusey2014anomalous}, using specific pre- and post-selected states.  We begin with a review of the Kochen-Specker (KS) theorem, and then discuss its interpretation using the Aharonov-Bergmann-Lebowitz (ABL)\footnote{We should emphasize that the ABL reformulation of quantum mechanics is completely consistent with the conventional mathematical formalism of quantum mechanics, and differs only in interpretation.} \cite{aharonov1964time} reformulation of quantum mechanics and particular pre- and post-selected states.  We next show that several classes of KS sets give rise to the quantum pigeonhole effect, which essentially denies the existence of classical physical states during the interval between pre- and post-selection.  Finally, we show that in such cases there is always a classical projector with a negative (anomalous) weak value that can be observed using weak measurements.

\subsubsection{The Kochen-Specker Theorem}
The Kochen-Specker theorem posits that noncontextual hidden-variable models of reality that include free random choice are inconsistent with quantum mechanics.  The KS theorem is proved by particular sets of observables and measurement contexts that are called {\it KS sets}.

\begin{figure}[h!]
\centering
\caption{The 3-qubit Peres-Mermin Square}
\includegraphics[width=3in]{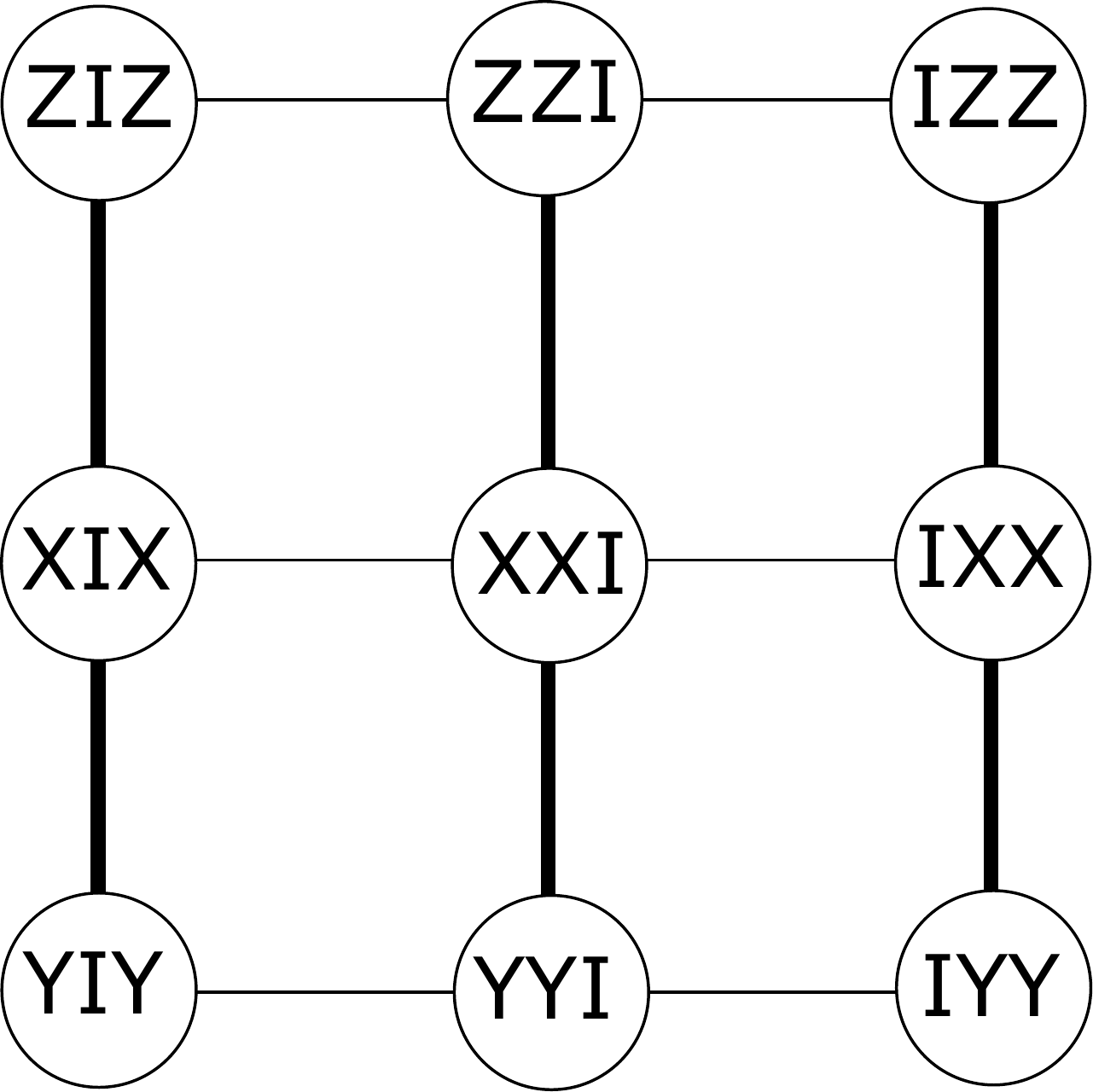}
\label{Square}
\end{figure}

A measurement {\it context} is any set of mutually commuting observables that can be jointly measured (this is also called a {\it basis}, and we will use the two terms interchangeably).  Any noncontextual hidden variable theory (NCHVT) requires that a truth-value (i.e. an eigenvalue) be preassigned to each observable within a set in such a way that the assigned value is independent of which contexts of mutually commuting observables the observable belongs to (i.e. the assignment is noncontextual).  If the noncontextual hidden variable theory is to agree with experiment, then the assignment must obey the `product rule' for any mutually commuting set of observables, and the `sum rule' for any set of mutually orthogonal projectors that sum to identity.  The product rule dictates that the product of the assigned eigenvalues in any context is equal to the eigenvalue assigned to the product of the observables in that context (Eq. \ref{Product}).  The sum rule dictates that if a set of mutually orthogonal projectors sum to identity, then the sum of the eigenvalues (0 or 1) assigned to that set must be 1 (Eq. \ref{Sum}).
\begin{equation}
AB=C \rightarrow \lambda_A\lambda_B=\lambda_c  \label{Product}
\end{equation}
\begin{equation}
\sum_i \Pi_i = I \rightarrow \sum_i \lambda_i = 1   \label{Sum}
\end{equation}
A KS set is any set of observables and contexts that cannot admit such an assignment, and thus such a set proves the KS theorem.  Specifically, if we insist on a NCHVT, then the sum rule or product rule must be violated in at least one context of the KS set.  For example, consider the 3-qubit KS set of observables and contexts shown in Fig. \ref{Square}.  Each observable in the set belongs to the 3-qubit Pauli group (the qubit indices and tensor products between single-qubit observables are usually omitted here and throughout the text for brevity - for example $ZIZ \equiv \hat{\sigma}_Z^1 \otimes \hat{I}^2 \otimes \hat{\sigma}_Z^3$).  The lines represent contexts of mutually commuting observables, and the product of all observables on a line is $+I^{\otimes 3}$ for thin lines, and $-I^{\otimes 3}$ for thick lines.  We call contexts with this property positive and negative {\it identity products} (IDs) respectively.  To see why this is a KS set, consider the overall product of all six IDs in the set.  In any NCHVT, we must assign a $\pm 1$ eigenvalue to each of the nine observables, but because each observable belongs to two contexts (a row and a column), all of these eigenvalues will appear squared in the overall product, which must then be +1.  However, if the product rule were to be obeyed in all six contexts, the product would be -1 since there are an odd number of negative IDs in the set, and thus the NCHVT must violate the product rule.

As a result of this violation, each KS set leads to a Bell-type inequality that can be used to experimentally test the KS theorem \cite{cabello2008experimentally, WA_600cell_2}, based on the assumption that randomly chosen measurements must necessarily reveal the context(s) where the violation occurs.  In Appendix \ref{KS_NQPauli} we give a review of KS sets from within the $N$-qubit Pauli group, and how they prove the KS theorem \cite{WA_24Rays, WA_60Rays, WA_3qubits, WA_4qubits, WA_Nqubits, WaegellThesis, W_Primitive}.

When we actually perform these experiments we find that the measurement outcomes never violate the sum or product rule, which shows that nature is inconsistent with NCHVTs.  In fact, the situation that the measurement outcome violates the sum or product rule cannot even be represented by a state in Hilbert space to begin with, and would thus be fundamentally inconsistent with quantum mechanics.  As a result, one cannot associate specific eigenvalues with the observables in contexts where such violations occur.  This is related to what has been called the `value-indefiniteness' of these observables \cite{abbott2012strong, Abbott2015}, as we discuss later.

More generally, a KS set is composed of some set of observables $\{O\}$ which form some set of contexts $\{C\}$ that cannot admit a complete noncontextual value assignment.  In the above example the observables are Pauli operators and the contexts are IDs, but this is just one special case.

\subsubsection{A Realist Model}

The KS theorem shows us that if we wish to preserve continuous ontological realism within quantum mechanics, we will have to give up on either the idea of free random choice, or the idea of noncontextuality.  We choose to adopt a realist hidden variable theory that still assigns a single truth-value, $v_e(O)$ (an eigenvalue) to each observable $O$ of the system, in such a way that the assigned value $v_e(O)$ is independent of which contexts $C$ of mutually commuting observables $O$ belongs to (which was our original definition of noncontextuality).  In making such an assignment to a KS set, the sum or product rule must still be violated, but now we shift the focus of the problem to explaining {\it why} the violation is never observed.  One explanation is a limitation on free choice, which is to say that nature somehow forbids us from choosing to observe the context(s) where the violation occurs.  Alternatively, the entire set of realist truth-values $v_e$ could explicitly agree with - and thus depend on - both the pre-selected state and the post-selected state, such that contexts containing these states have definite truth-value assignments that obey the rules, and any violations are shifted into other contexts.  These two interpretations are operationally equivalent, in that they both assume that violations exist but are never observed - in the former view it is the preexisting truth-value assignment that limits future free choice, while in the latter it is future free choice that limits the preexisting truth-value assignment.  Although the latter view might seem to challenge some notions of causality, the relative inaccessibility of these truth values preserves no-signalling.  For what follows, we are free to remain agnostic as to which of these two interpretations to adopt, since we are only concerned with the realist truth-value assignment, $v_e$, itself.

In this scenario we have chosen a pre- and post-selection (PPS) from within the projectors of the KS set.  We then can call the explicit dependence of $v_e$ on the pre- and post-selected states {\it PPS-contextuality}.  It is noteworthy that weak values are PPS-contextual in this way by definition.  It is also important to note that all known PPS paradoxes \cite{aharonov2010time, cabello1997no, aharonov2014quantum, leifer2005pre}, such as the 3-box paradox, are based implicitly on a hidden variable model of the type $v_e$, and in these cases the paradox {\it is} the violation of the sum or product rule.  The subtle point is that all of these cases take $v_e$ for granted in order to obtain the paradox, and thus they must implicitly obey one of the interpretations given above regarding free choice or retrocausality.  It is therefore dubious to think of $v_e$ as a {\it classical} hidden variable model - it is more accurately a PPS-contextual hidden variable model, with some aspects of noncontextuality left intact.  This observation is an important generalization of the work of Leifer and Spekkens \cite{leifer2005pre}.

In a given experiment, we can learn a subset of the realist truth-values $v_e(O)$ assigned to the observables of the system: naturally we learn the values assigned to the observables in the context that we actually measure.  We can also choose to prepare a specific eigenstate of another context in the KS set rather than starting with an arbitrary state, and thus we also know the realist truth-values in that context.  There is no experiment in which we can directly learn the assignments $v_e(O)$ to additional observables, and the physical meaning of such values is unclear.

\subsubsection{The ABL Reformulation, Weak Values, and the Quantum Pigeonhole Effect }

Using the ABL reformulation of quantum mechanics, we can indirectly extend the realist truth-value assignment by using weak values for the specific pre- and post-selection (see Appendix \ref{ABLDerivation} for a brief review of the machinery of ABL).  In some situations, this extended view of the definite eigenvalue assignment allows us to identify exactly in which context(s) the sum or product rule is violated, which we call {\it conflict bases}.  In some interpretations of the ABL reformulation, the weak values of observables $v_w(O)$ are treated as physical properties of the system during the interval between the pre- and post-selection.  Indeed, $v_w(O)$ is automatically guaranteed to agree with $v_e(O)$ for all observables whose eigenvalues are fixed by the PPS, but more importantly, the weak values give us an alternate valuation of all of the observables in the set.  The weak values are not generally constrained by the eigenvalue spectrum of the observables, and we call weak values outside the range of this spectrum `anomalous.'  There are however particular cases where the weak value reveals the eigenvalue that a strong measurement would have obtained, and we interpret these cases as revealing additional realist truth-values as well.

The weak values of projectors $v_w(\Pi_i)$ are independent of which bases they appear in (our original notion of noncontextuality), but unlike the 0s and 1s of realist truth-value assignments $v_e(\Pi_i)$, they always obey the sum rule as a result of their definition.  They are also explicitly PPS-contextual in the sense that they are only defined for particular pre- and post-selected states (call them $|\Psi\rangle$ and $|\Phi\rangle$ respectively),
\begin{equation}
v_w(\Pi_i) = \frac{\langle\Phi|\Pi_i|\Psi\rangle}{\langle\Phi|\Psi\rangle}.
\end{equation}

We can weakly measure the projectors in the specific bases where the realist truth-value assignment $v_e$ violates the sum rule, and these measurements reveal a specific signature.  In particular, there is always at least one conflict basis in which we find projectors with negative weak values.  This shows that proofs of the KS theorem can give rise to anomalous weak values in the ABL formulation, which lends some framework to Pusey's recent result \cite{pusey2014anomalous} showing that any anomalous weak value is a proof of measurement contextuality.  Furthermore, using the ABL formula, it is also true that in all conflict bases in a KS set, all of the projectors that do not have zero probability instead have equal probability, which gives an operational meaning to the notion of value-indefiniteness - each eigenvalue is equally likely during the interval between pre- and post-selection.  This is the most general experimental signature of KS contextuality.

Finally, if the violation occurs in a `classical basis,' then we have an example of the quantum pigeonhole principle\footnote{In brief, the classical pigeonhole principle states that if $N$ objects are placed in $M<N$ separate boxes, then at least one box must contain more than one object.  In \cite{aharonov2014quantum} it is argued that quantum systems do not obey this principle.} \cite{aharonov2014quantum}, where a `classical basis' is intuitively defined as a basis composed of product states without superposition.  Specifically, for 3 pigeons in 2 holes, these are the states like, $|$left,right,right$\rangle$, or $|$left,left,left$\rangle$, etc$\ldots$.  We typically take this to be the product Pauli $Z$ basis, $\{|Z=+1\rangle \equiv|$left$\rangle$, $|Z=-1\rangle \equiv |$right$\rangle\}$.  The violation of the product rule for a KS set of observables and contexts, or the sum rule for a KS set of projectors and bases, in the classical basis, shows that the system cannot be in any of the classical states, and thus it is impossible to specify that any pigeon is in any specific hole.

\subsubsection{Organization}

The remainder of this paper is organized as follows: in the next section we explore the details of KS sets with pre- and post-selection using the truth-value assignment in our realist model alongside the weak values of the observables in the set.  Following this, we use the tools we have developed to give a detailed analysis of the 3-qubit KS set of Fig. \ref{Square}.  The main paper concludes and we give extensive theoretical development in the appendices of the ABL formulation with projectors of arbitrary rank, the $N$-qubit Pauli group and KS sets it contains, and several examples of KS sets and how they give rise to anomalous weak values, including a family that shows the pigeonhole effect for all $N$.  In Appendix \ref{MeanKing} we use the machinery we have introduced and weak values to show how any KS set can be used to guarantee that the Mean King wins the game, generalizing a result from Mermin \cite{mermin1995limits}.


\subsection{Realist Truth-Value Assignments and Weak Values}

As discussed in the introduction, by using pre- and post-selections that are already contained within a KS set, we learn part of the truth-value assignment $v_e$ that our realist model must make to that same KS set of observables.  In addition, we can consider the weak values $v_w$ of all of the observables in the set, and for observables where the weak value, $v_w(O)$, is also an eigenvalue, we reveal additional realist truth-value assignments, $v_e(O)$.  However, the weak values are not generally constrained to the eigenvalue spectrum of the observables, and so we cannot generally reveal a complete assignment of truth-values in this way.  The most interesting results of this paper are obtained by comparing the complete realist truth-value assignment, $v_e$, to a KS set with the complete weak valuation $v_w$ of that same set.

The ABL formula gives us the probability to obtain a particular outcome for a strong measurement on some observable that is made in the intermediate time between the pre- and post-selection \cite{aharonov1964time}.  It is a theorem that if the ABL probability to obtain a particular eigenvalue is 1, then the weak value of this observable, obtained by an intermediate weak measurement, is equal to that eigenvalue.  If such an eigenvalue does not come directly from the PPS, we call it a {\it forced value.}  This is widely known as the {\it ABL rule} \cite{tollaksen2007pre} (see Appendix \ref{ABLDerivation} for the derivation).

If a strong measurement of some observable had been performed during the intermediate time, it would have disturbed the PPS, and we would consequently alter the entire realist truth-value assignment, $v_e$, but the weak value $v_w(O)$ of that observable exists during the intermediate time even if we do not measure it.  The {\it reverse ABL rule} furthermore shows that if the weak value of a projector is equal to 1, then the ABL probability to obtain that projector by an intermediate strong measurement is 1 \cite{aharonov1991complete}.  Thus, if we further postulate that ideal weak measurements do not disturb the PPS or corresponding realist truth-value assignment $v_e$, then weak measurements of weak values can be used to justify inclusion of the forced eigenvalues in our realist truth-value assignment.

Pusey and Leifer \cite{pusey2015logical} have given an alternative argument that shows that if any projector has ABL probability 1 given a particular PPS, then that projector must also be assigned a 1 in any noncontextual ontological model (strictly speaking, a measurement noncontextual model with outcome determinism for sharp measurements) - independent of any weak measurements to probe that value.

For KS sets composed of Pauli observables and IDs, this happens in any ID for which all but one observable has an eigenvalue fixed by the PPS, as shown in Fig. \ref{SquareSteps} of the next section.  Under these circumstances, the ABL rule dictates that the weak value of this one observable is forced to match the sign required by the ID, effectively revealing another realist truth-value.  The values forced by several different IDs may then violate the product rule for another ID, demonstrating quantum contextuality.

For KS sets composed of projectors and bases, a forced value occurs for an unassigned projector in any basis in which all other projectors are assigned a value 0 by the PPS.  Under these circumstances, the weak value of this projector must be 1 in order to satisfy the sum rule, and we can include this forced value in our truth-value assignment $v_e$, as shown by the underlined `\underline{1}'s in rows 4-6 of Figs. \ref{KS} and \ref{BasesColor} for the example case of the next section.  If we wish to continue to be faithful to the sum rule within our partial truth-value assignment, we must then also force all of the projectors orthogonal to that one to have truth-value 0, as shown by the underlined `\underline{0}'s in Figs. \ref{KS} and \ref{BasesColor}.  It is important to note that the weak values of these projectors need not also be 0 (which can be seen by contrasting the top rows of Figs. \ref{Weak} and \ref{KS}).

By this stage, our realist truth-value assignment $v_e$ may contain a violation of the sum rule for one or more bases within the set.  We may find a basis with all of the projectors assigned the value 0, or we may find one with more than one projector assigned the value 1.  In this way, the extension of $v_e$ provided by the ABL formulation, weak values $v_w$, and weak measurements allows us to identify the specific bases in which the conflict occurs, and for the classical basis, this is the quantum pigeonhole effect.

Beginning in Appendix \ref{Wheel3Derivation} we derive the general signatures of weak values in conflict bases, as discussed in the introduction.

\subsection{The Quantum Pigeonhole Effect}

As we have described above, the quantum pigeonhole effect arises as a special state-dependent case of quantum contextuality, in which we find the logical conflict, or value-indefiniteness, in the classical basis.  In this section, we will work through an example in detail in order to ensure that the key concepts previously introduced are conveyed clearly.  Additional examples are examined in Appendices \ref{Wheel3Derivation}, \ref{Wheel4Derivation}, and \ref{ArchDerivation}.

We use a particular notation that avoids the usual Hilbert space vector representation, and instead represents the state as a list of stabilizer observables and corresponding eigenvalues.  The stabilizer group is the complete group of mutually commuting $N$-qubit Pauli observables $\{O_i\}$ with the specified state as a joint eigenstate, and with the specified eigenvalues $\{\lambda_i\}$ (with $\lambda_i= \pm1$).  This eigenstate is actually a projector of rank $r = 2^{N-l}$, where $l$ is the number of independent observables in the set \footnote{In linear algebra, the action of projector of rank-$r$ is to project any object represented in the original $d$-dimensional vector space into a specific $r$-dimensional subspace.  The usual quantum measurement is a rank-1 projector that projects onto a specific vector in the Hilbert space - where a vector is a 1-dimensional subspace.  We can then think of a rank-$r$ projector in quantum mechanics as representing an `incomplete' collapse of the wavefunction.}.  For example, we represent an eigen-projector of three arbitrary mutually commuting observables $A$, $B$, and $C$, with corresponding eigenvalues $\lambda_A$, $\lambda_B$, and $\lambda_C$ as simply,
\begin{equation}
|\lambda_A A, \lambda_B B, \lambda_C C|.
\end{equation}
A ket of the form,
\begin{equation}
|\lambda_A A, \lambda_B B, \lambda_C C\rangle,
\end{equation}
indicates any state (i.e. its outer product with itself is a rank-1 projector) within the $r$-dimensional subspace of the projector.  Therefore, we can expand the rank-$r$ projector using kets as,
\begin{equation}
|\lambda_A A, \lambda_B B, \lambda_C C|
\end{equation}
\begin{equation}
=\sum_i^r |\lambda_A A, \lambda_B B, \lambda_C C, \ldots \rangle_i \langle\lambda_A A, \lambda_B B, \lambda_C C,\ldots|_i,
\end{equation}
where $\{|\lambda_A A, \lambda_B B, \lambda_C C, \ldots \rangle_i\}$ is any orthonormal basis that spans the projector subspace.

\subsubsection{The Pauli Observables and IDs Picture}

\begin{figure}[h!]
\caption{(Color Online) The 3-qubit Peres-Mermin Square with a particular PPS gives rise to a violation of the product rule in the classical basis, and thus a demonstration of the quantum pigeonhole effect.}
\centering
\subfloat[][No assignment]{
\includegraphics[width=1.25in]{Mermin3Square.pdf}}
\qquad
\subfloat[][Pre- and post-selection]{
\includegraphics[width=1.25in]{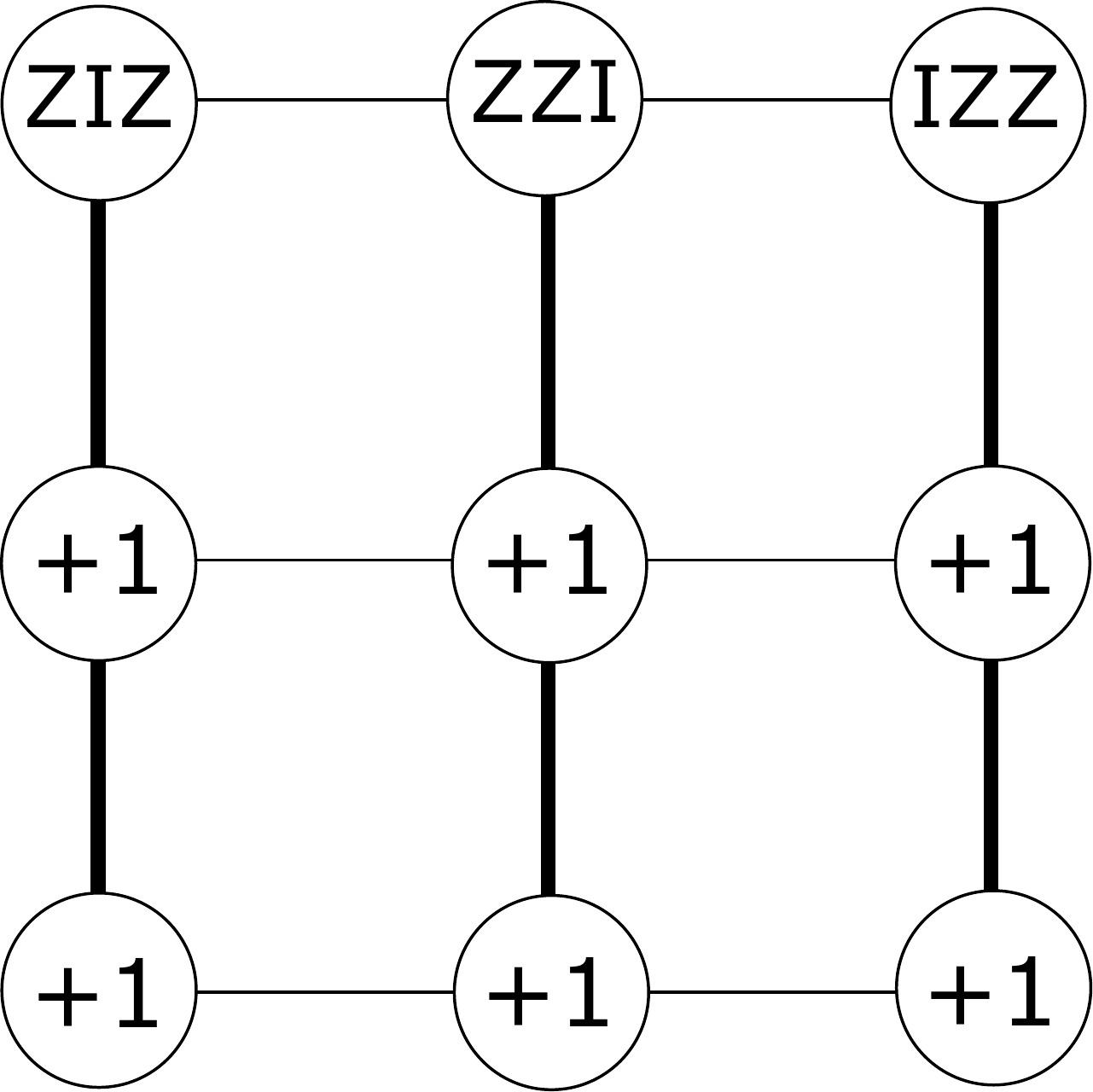}}
\qquad
\subfloat[][The ABL rule and violation of the product rule (shown by the blue dashed line)]{
\includegraphics[width=1.25in]{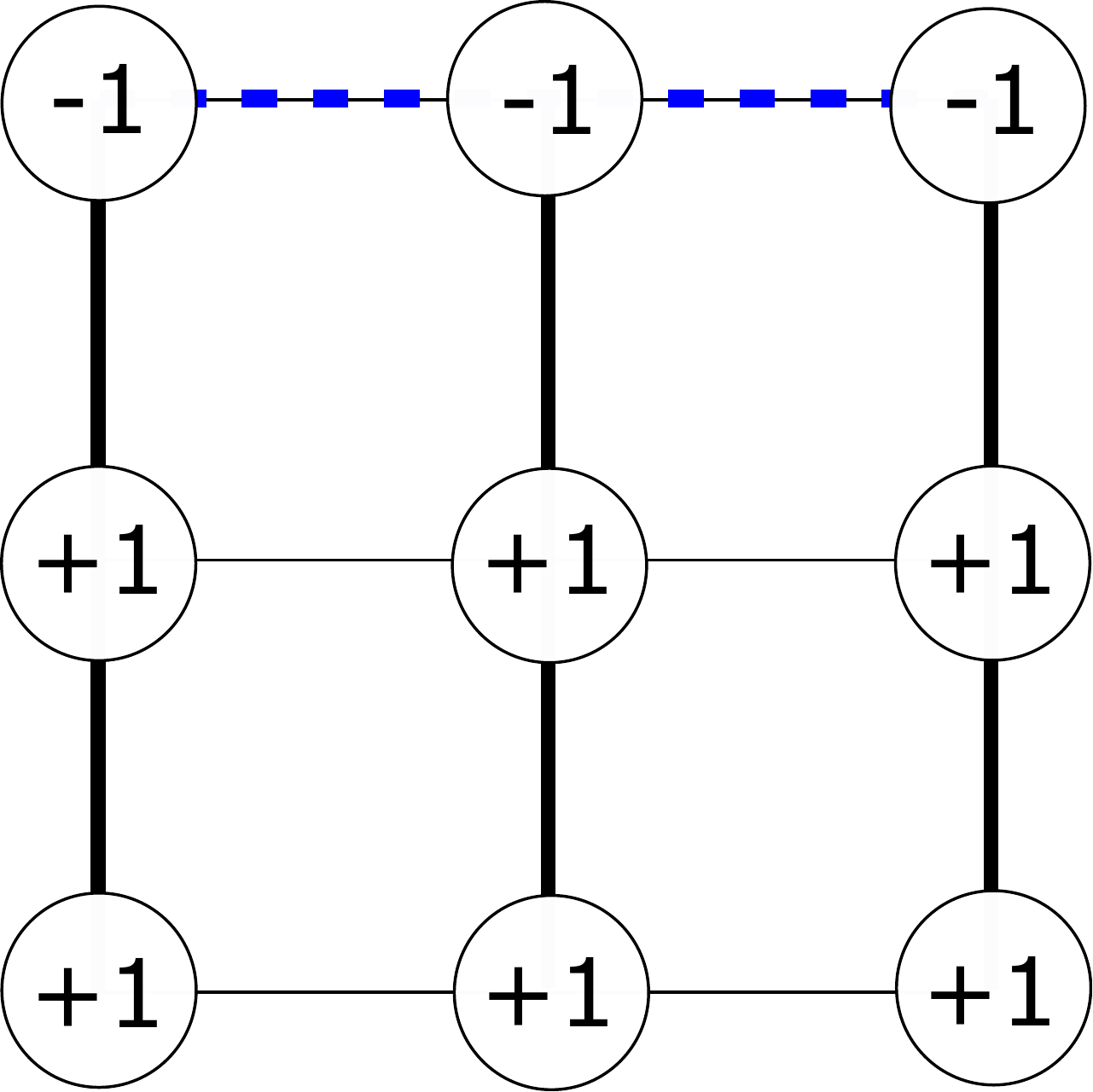}\label{SquareS3}}\label{SquareSteps}
\end{figure}

We begin with the KS set of Fig. \ref{Square}, and we choose as our pre-selection the product state $|\Psi\rangle = |+X_1,+X_2,+X_3\rangle$ and as our post-selection the product state $|\Phi\rangle = |+Y_1,+Y_2,+Y_3\rangle$.  This PPS fixes the eigenvalues of the observables on the two lower horizontal IDs to be +1.  Now, because each vertical ID is negative (as indicated by the bold vertical lines) and each has only one observable that is not assigned a value by this PPS, the ABL rule gives unit probability for the eigenvalue of the last observable to be -1 (i.e. along the top row).  These are examples of forced values, and they introduce a conflict in the top horizontal ID, because they violate the product rule for that ID (which is positive, as indicated by the narrow horizontal line).  The steps of this process are shown in Fig. \ref{SquareSteps}.  The top row of Fig \ref{Square} is a conflict ID and has the classical basis as its joint eigenbasis, and thus we have demonstrated the quantum pigeonhole effect: with this KS set and this PPS, none of the classical states can exist during the intermediate time.  However, as we will show, we {\it can} learn the weak values of the projectors onto the classical states through weak measurements, and these weak values display a particular signature.

\subsubsection{The Stabilizer Projector Picture}

It will be informative to recast this example in terms of the projectors.  For compactness, we will introduce alphabetical labels $A,\ldots,I$ for the observables of Fig. \ref{Square}, proceeding from left to right and top to bottom.  Thus, $A \equiv ZIZ$, $B\equiv ZZI$, $\ldots$, $I \equiv IYY$ (the distinction between these two uses of $I$ should be obvious from their usage).  The pre-selection $|\Psi\rangle$ and post-selection $|\Phi\rangle$ can be expressed by the kets,
\begin{equation}
|\Psi\rangle = |+D, +E, +F\rangle
\end{equation}
\begin{equation}
|\Phi\rangle = |+G, +H, +I\rangle,
\end{equation}
where the $+$ or $-$ signs indicate the eigenvalue of the specified observable in the given PPS state.

From the ABL rule, we can use this PPS, and the fact that each vertical ID in Fig. \ref{Square} is negative, to force several other rank-2 projectors.  These are
\begin{equation}
|f_1| = |-A, +D, +G|, \label{f1}
\end{equation}
\begin{equation}
|f_2| = |-B, +E, +H|,  \label{f2}
\end{equation}
and
\begin{equation}
|f_3| = |-C, +F, +I|.  \label{f3}
\end{equation}
The ABL rule gives unit probability to obtain these states if a strong measurement is performed during the intermediate time, and therefore, the weak values of the projectors onto these states are 1, and vice versa (by the reverse ABL rule).  As described above, we can use ideal weak measurements of these projectors to justify including these forced values in our realist truth-value assignment $v_e$.  The next step is to assign forced 0s to all of the states orthogonal to those with forced 1s, and this is the step that gives us the pigeonhole effect.

The classical basis is composed of the rank-2 projectors,
\begin{equation}
\{|+A, +B, +C|,|+A, -B, -C|,|-A, +B, -C|,|-A, -B, +C|\}.\label{ABC}
\end{equation}
Two projectors are orthogonal if they are both eigen-projectors of the same observable, but with different eigenvalues, and therefore, all four projectors in the classical basis will be assigned the value 0 in our realist truth-value assignment $v_e$, because they are all orthogonal to at least one of the forced states of Eqs. \ref{f1},\ref{f2}, and \ref{f3}, and thus the violation of the sum rule corresponds exactly to the pigeonhole effect.  It is also worth noting that $|+A, +B, +C|$ is orthogonal to all three of the forced projectors, and as we will see, this is also the projector with an anomalous weak value.  We can think of this as the {\it maximum conflict projector}, and as we show for a variety of other KS sets in Appendices \ref{Wheel3Derivation}, \ref{Wheel4Derivation}, and \ref{ArchDerivation}, the maximum conflict projector always has an anomalous weak value.

\begin{figure}[h!]
\caption{}
\centering
\subfloat[][Each row shows one of the 24 orthogonal bases, with the rank-2 projectors indexed 1 through 24.  The first 3 bases are the eigenbases of the horizontal IDs of Fig. \ref{Square} and the next 3 bases are the eigenbases of the vertical IDs.  The other 18 are hybrid bases.]{
\begin{tabular}{|cccc|}
\hline
1 & 2 & 3 & 4 \\
5 & 6 & 7 & 8 \\
9 & 10 & 11 & 12 \\
13 & 14 & 15 & 16 \\
17 & 18 & 19 & 20 \\
21 & 22 & 23 & 24 \\
\hline
1 & 2 & 13 & 16 \\
1 & 3 & 17 & 20 \\
1 & 4 & 21 & 24 \\
2 & 3 & 22 & 23 \\
2 & 4 & 18 & 19 \\
3 & 4 & 14 & 15 \\
5 & 6 & 15 & 16 \\
5 & 7 & 19 & 20 \\
5 & 8 & 23 & 24 \\
6 & 7 & 21 & 22 \\
6 & 8 & 17 & 18 \\
7 & 8 & 13 & 14 \\
9 & 10 & 14 & 16 \\
9 & 11 & 18 & 20 \\
9 & 12 & 22 & 24 \\
10 & 11 & 21 & 23 \\
10 & 12 & 17 & 19 \\
11 & 12 & 13 & 15 \\
\hline
\end{tabular}\label{Rays}}
\qquad
\subfloat[][The weak values $v_w$ of the 24 projectors given the PPS $|\Psi\rangle=|+X_1,+X_2,+X_3\rangle$ and $|\Phi\rangle=|+Y_1,+Y_2,+Y_3\rangle$, which sets projectors 5 and 9 to truth-value 1, and orthogonal projectors to 0.  The weak value of each projector $v(\Pi_i)_w$ is shown superimposed on that projector's location in \subref{Rays}.]{
\begin{tabular}{|cccc|}
\hline
-0.5 & 0.5 & 0.5 & 0.5 \\
1 & 0 & 0 & 0 \\
1 & 0 & 0 & 0 \\
1 & 0 & 0 & 0 \\
1 & 0 & 0 & 0 \\
1 & 0 & 0 & 0 \\
\hline
-0.5 & 0.5 & 1 & 0 \\
-0.5 & 0.5 & 1 & 0 \\
-0.5 & 0.5 & 1 & 0 \\
0.5 & 0.5 & 0 & 0 \\
0.5 & 0.5 & 0 & 0 \\
0.5 & 0.5 & 0 & 0 \\
1 & 0 & 0 & 0 \\
1 & 0 & 0 & 0 \\
1 & 0 & 0 & 0 \\
0 & 0 & 1 & 0 \\
0 & 0 & 1 & 0 \\
0 & 0 & 1 & 0 \\
1 & 0 & 0 & 0 \\
1 & 0 & 0 & 0 \\
1 & 0 & 0 & 0 \\
0 & 0 & 1 & 0 \\
0 & 0 & 1 & 0 \\
0 & 0 & 1 & 0 \\
\hline
\end{tabular}\label{Weak}}
\qquad
\subfloat[][The realist truth-value assignment $v_e$ for this PPS, with underlines for the forced 1s and 0s.  The realist truth-value of each projector $v(\Pi_i)_e$ is shown superimposed on that projector's location in \subref{Rays}.  Conflict bases are shown in bold, and the first basis is the classical basis.]{
\begin{tabular}{|cccc|}
\hline
\textbf{\underline{0}} & \textbf{\underline{0}} & \textbf{\underline{0}} & \textbf{\underline{0}} \\
1 & 0 & 0 & 0 \\
1 & 0 & 0 & 0 \\
\underline{1} & 0 & 0 & 0 \\
\underline{1} & 0 & 0 & 0 \\
\underline{1} & 0 & 0 & 0 \\
\hline
\underline{0} & \underline{0} & \underline{1} & 0 \\
\underline{0} & \underline{0} & \underline{1} & 0 \\
\underline{0} & \underline{0} & \underline{1} & 0 \\
\textbf{\underline{0}} & \textbf{\underline{0}} & \textbf{0} & \textbf{0} \\
\textbf{\underline{0}} & \textbf{\underline{0}} & \textbf{0} & \textbf{0} \\
\textbf{\underline{0}} & \textbf{\underline{0}} & \textbf{0} & \textbf{0} \\
1 & 0 & 0 & 0 \\
1 & 0 & 0 & 0 \\
1 & 0 & 0 & 0 \\
0 & 0 & \underline{1} & 0 \\
0 & 0 & \underline{1} & 0 \\
0 & 0 & \underline{1} & 0 \\
1 & 0 & 0 & 0 \\
1 & 0 & 0 & 0 \\
1 & 0 & 0 & 0 \\
0 & 0 & \underline{1} & 0 \\
0 & 0 & \underline{1} & 0 \\
0 & 0 & \underline{1} & 0 \\
\hline
\end{tabular}\label{KS}}
\label{Bases}
\end{figure}

\begin{figure}[h!]
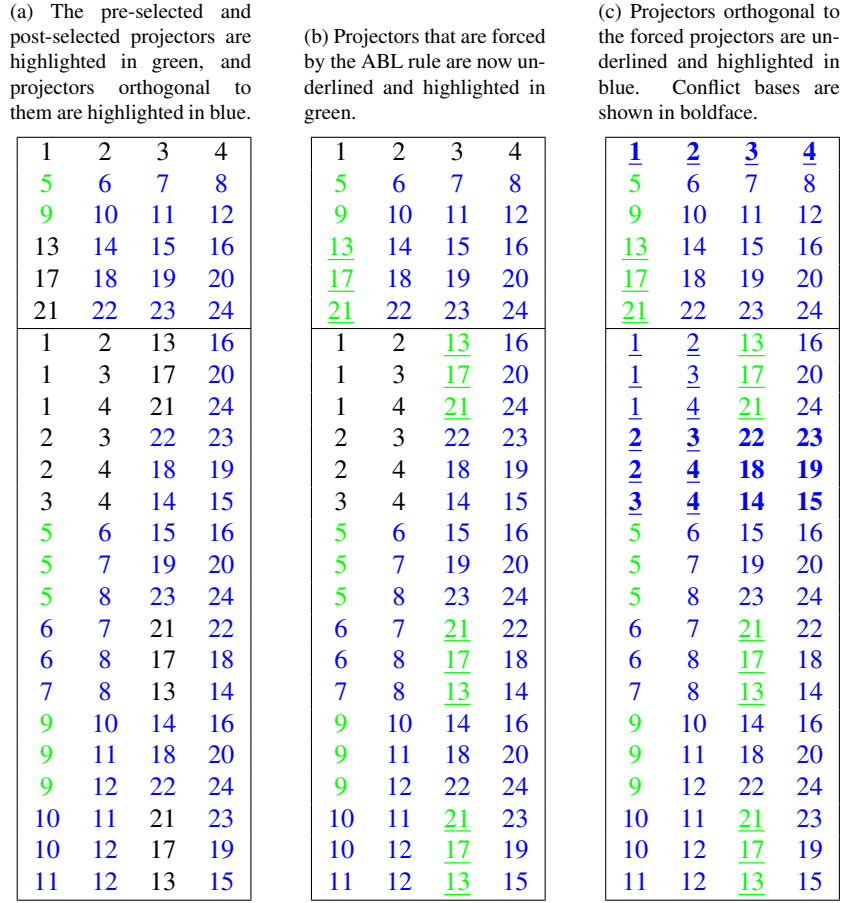

\caption{(Color Online) Realist truth-value ($v_e$) assignment process for the 24 projectors.}
\centering
\subfloat[][The pre-selected and post-selected projectors are highlighted in green, and projectors orthogonal to them are highlighted in blue.]{
\begin{tabular}{|cccc|}
\hline
1 & 2 & 3 & 4 \\
\textcolor{green}{5} & \textcolor{blue}{6} & \textcolor{blue}{7} & \textcolor{blue}{8} \\
\textcolor{green}{9} & \textcolor{blue}{10} & \textcolor{blue}{11} & \textcolor{blue}{12} \\
13 & \textcolor{blue}{14} & \textcolor{blue}{15} & \textcolor{blue}{16} \\
17 & \textcolor{blue}{18} & \textcolor{blue}{19} & \textcolor{blue}{20} \\
21 & \textcolor{blue}{22} & \textcolor{blue}{23} & \textcolor{blue}{24} \\
\hline
1 & 2 & 13 & \textcolor{blue}{16} \\
1 & 3 & 17 & \textcolor{blue}{20} \\
1 & 4 & 21 & \textcolor{blue}{24} \\
2 & 3 & \textcolor{blue}{22} & \textcolor{blue}{23} \\
2 & 4 & \textcolor{blue}{18} & \textcolor{blue}{19} \\
3 & 4 & \textcolor{blue}{14} & \textcolor{blue}{15} \\
\textcolor{green}{5} & \textcolor{blue}{6} & \textcolor{blue}{15} & \textcolor{blue}{16} \\
\textcolor{green}{5} & \textcolor{blue}{7} & \textcolor{blue}{19} & \textcolor{blue}{20} \\
\textcolor{green}{5} & \textcolor{blue}{8} & \textcolor{blue}{23} & \textcolor{blue}{24} \\
\textcolor{blue}{6} & \textcolor{blue}{7} & 21 & \textcolor{blue}{22} \\
\textcolor{blue}{6} & \textcolor{blue}{8} & 17 & \textcolor{blue}{18} \\
\textcolor{blue}{7} & \textcolor{blue}{8} & 13 & \textcolor{blue}{14} \\
\textcolor{green}{9} & \textcolor{blue}{10} & \textcolor{blue}{14} & \textcolor{blue}{16} \\
\textcolor{green}{9} & \textcolor{blue}{11} & \textcolor{blue}{18} & \textcolor{blue}{20} \\
\textcolor{green}{9} & \textcolor{blue}{12} & \textcolor{blue}{22} & \textcolor{blue}{24} \\
\textcolor{blue}{10} & \textcolor{blue}{11} & 21 & \textcolor{blue}{23} \\
\textcolor{blue}{10} & \textcolor{blue}{12} & 17 & \textcolor{blue}{19} \\
\textcolor{blue}{11} & \textcolor{blue}{12} & 13 & \textcolor{blue}{15} \\
\hline
\end{tabular}\label{S1}}
\qquad
\subfloat[][Projectors that are forced by the ABL rule are now underlined and highlighted in green.]{
\begin{tabular}{|cccc|}
\hline
1 & 2 & 3 & 4 \\
\textcolor{green}{5} & \textcolor{blue}{6} & \textcolor{blue}{7} & \textcolor{blue}{8} \\
\textcolor{green}{9} & \textcolor{blue}{10} & \textcolor{blue}{11} & \textcolor{blue}{12} \\
\textcolor{green}{\underline{13}} & \textcolor{blue}{14} & \textcolor{blue}{15} & \textcolor{blue}{16} \\
\textcolor{green}{\underline{17}} & \textcolor{blue}{18} & \textcolor{blue}{19} & \textcolor{blue}{20} \\
\textcolor{green}{\underline{21}} & \textcolor{blue}{22} & \textcolor{blue}{23} & \textcolor{blue}{24} \\
\hline
1 & 2 & \textcolor{green}{\underline{13}} & \textcolor{blue}{16} \\
1 & 3 & \textcolor{green}{\underline{17}} & \textcolor{blue}{20} \\
1 & 4 & \textcolor{green}{\underline{21}} & \textcolor{blue}{24} \\
2 & 3 & \textcolor{blue}{22} & \textcolor{blue}{23} \\
2 & 4 & \textcolor{blue}{18} & \textcolor{blue}{19} \\
3 & 4 & \textcolor{blue}{14} & \textcolor{blue}{15} \\
\textcolor{green}{5} & \textcolor{blue}{6} & \textcolor{blue}{15} & \textcolor{blue}{16} \\
\textcolor{green}{5} & \textcolor{blue}{7} & \textcolor{blue}{19} & \textcolor{blue}{20} \\
\textcolor{green}{5} & \textcolor{blue}{8} & \textcolor{blue}{23} & \textcolor{blue}{24} \\
\textcolor{blue}{6} & \textcolor{blue}{7} & \textcolor{green}{\underline{21}} & \textcolor{blue}{22} \\
\textcolor{blue}{6} & \textcolor{blue}{8} & \textcolor{green}{\underline{17}} & \textcolor{blue}{18} \\
\textcolor{blue}{7} & \textcolor{blue}{8} & \textcolor{green}{\underline{13}} & \textcolor{blue}{14} \\
\textcolor{green}{9} & \textcolor{blue}{10} & \textcolor{blue}{14} & \textcolor{blue}{16} \\
\textcolor{green}{9} & \textcolor{blue}{11} & \textcolor{blue}{18} & \textcolor{blue}{20} \\
\textcolor{green}{9} & \textcolor{blue}{12} & \textcolor{blue}{22} & \textcolor{blue}{24} \\
\textcolor{blue}{10} & \textcolor{blue}{11} & \textcolor{green}{\underline{21}} & \textcolor{blue}{23} \\
\textcolor{blue}{10} & \textcolor{blue}{12} & \textcolor{green}{\underline{17}} & \textcolor{blue}{19} \\
\textcolor{blue}{11} & \textcolor{blue}{12} & \textcolor{green}{\underline{13}} & \textcolor{blue}{15} \\
\hline
\end{tabular}\label{S2}}
\qquad
\subfloat[][Projectors orthogonal to the forced projectors are underlined and highlighted in blue.  Conflict bases are shown in boldface.]{
\begin{tabular}{|cccc|}
\hline
\textbf{\textcolor{blue}{\underline{1}}} & \textbf{\textcolor{blue}{\underline{2}}} & \textbf{\textcolor{blue}{\underline{3}}} & \textbf{\textcolor{blue}{\underline{4}}} \\
\textcolor{green}{5} & \textcolor{blue}{6} & \textcolor{blue}{7} & \textcolor{blue}{8} \\
\textcolor{green}{9} & \textcolor{blue}{10} & \textcolor{blue}{11} & \textcolor{blue}{12} \\
\textcolor{green}{\underline{13}} & \textcolor{blue}{14} & \textcolor{blue}{15} & \textcolor{blue}{16} \\
\textcolor{green}{\underline{17}} & \textcolor{blue}{18} & \textcolor{blue}{19} & \textcolor{blue}{20} \\
\textcolor{green}{\underline{21}} & \textcolor{blue}{22} & \textcolor{blue}{23} & \textcolor{blue}{24} \\
\hline
\textcolor{blue}{\underline{1}} & \textcolor{blue}{\underline{2}} & \textcolor{green}{\underline{13}} & \textcolor{blue}{16} \\
\textcolor{blue}{\underline{1}} & \textcolor{blue}{\underline{3}} & \textcolor{green}{\underline{17}} & \textcolor{blue}{20} \\
\textcolor{blue}{\underline{1}} & \textcolor{blue}{\underline{4}} & \textcolor{green}{\underline{21}} & \textcolor{blue}{24} \\
\textbf{\textcolor{blue}{\underline{2}}} & \textbf{\textcolor{blue}{\underline{3}}} & \textbf{\textcolor{blue}{22}} & \textbf{\textcolor{blue}{23}} \\
\textbf{\textcolor{blue}{\underline{2}}} & \textbf{\textcolor{blue}{\underline{4}}} & \textbf{\textcolor{blue}{18}} & \textbf{\textcolor{blue}{19}} \\
\textbf{\textcolor{blue}{\underline{3}}} & \textbf{\textcolor{blue}{\underline{4}}} & \textbf{\textcolor{blue}{14}} & \textbf{\textcolor{blue}{15}} \\
\textcolor{green}{5} & \textcolor{blue}{6} & \textcolor{blue}{15} & \textcolor{blue}{16} \\
\textcolor{green}{5} & \textcolor{blue}{7} & \textcolor{blue}{19} & \textcolor{blue}{20} \\
\textcolor{green}{5} & \textcolor{blue}{8} & \textcolor{blue}{23} & \textcolor{blue}{24} \\
\textcolor{blue}{6} & \textcolor{blue}{7} & \textcolor{green}{\underline{21}} & \textcolor{blue}{22} \\
\textcolor{blue}{6} & \textcolor{blue}{8} & \textcolor{green}{\underline{17}} & \textcolor{blue}{18} \\
\textcolor{blue}{7} & \textcolor{blue}{8} & \textcolor{green}{\underline{13}} & \textcolor{blue}{14} \\
\textcolor{green}{9} & \textcolor{blue}{10} & \textcolor{blue}{14} & \textcolor{blue}{16} \\
\textcolor{green}{9} & \textcolor{blue}{11} & \textcolor{blue}{18} & \textcolor{blue}{20} \\
\textcolor{green}{9} & \textcolor{blue}{12} & \textcolor{blue}{22} & \textcolor{blue}{24} \\
\textcolor{blue}{10} & \textcolor{blue}{11} & \textcolor{green}{\underline{21}} & \textcolor{blue}{23} \\
\textcolor{blue}{10} & \textcolor{blue}{12} & \textcolor{green}{\underline{17}} & \textcolor{blue}{19} \\
\textcolor{blue}{11} & \textcolor{blue}{12} & \textcolor{green}{\underline{13}} & \textcolor{blue}{15} \\
\hline
\end{tabular}\label{S3}}
\label{BasesColor}
\end{figure}

It is important to note that this last step of forcing 0 assignments is also where the realist truth-value assignments $v_e$ begin to diverge from the weak values $v_w$ of the projectors.  The weak values must obey the sum rule, and thus they clearly cannot all be 0 in a given basis.  In fact, the actual weak values of the projectors $v(\Pi)$ that appear in the basis where our realist truth-value assignment $v_e$ violates the sum rule are the only experimental access we have to this logical conflict.

The rows of Fig. \ref{Bases} show the 24 orthogonal bases formed by these 24 rank-2 projectors, the corresponding weak values given this PPS, and the realist truth-value assignment.  This particular set of projectors and bases is {\it saturated}, meaning that two projectors are orthogonal if and only if they appear together in at least one basis.  This makes the basis table completely equivalent to the {\it Kochen-Specker diagram} for this set of projectors.  The KS diagram is a graph with a vertex corresponding to each projector, and edges connecting vertices that correspond to orthogonal pairs of projectors\footnote{The set we are discussing here is isomorphic to the set of 24 rank-1 projectors and 24 orthogonal bases that arise from the 2-qubit Peres-Mermin Square \cite{WA_24Rays}, meaning that the pattern of orthogonalities between projectors is identical in both cases (although the indexing is not exactly the same between the two papers).  The set is represented by 24 pure state vectors in a 4-dimensional Hilbert space, which may be a good starting point for building an intuition about these structures.}.

The four projectors in each of the six eigenbasis are indexed in ascending order of their eigenvalues for the first two observables in each ID, in the order $(++,+-,-+,--)$.  For example, Eq. \ref{ABC} shows projectors 1-4, the joint eigenbasis of observables $A \equiv ZIZ$ and $B \equiv ZZI$).  Furthermore, the rank-2 projector indexed 5 contains the pre-selected state $|+X_1,+X_2,+X_3\rangle$, and projector 9 contains the post-selected state $|+Y_1,+Y_2,+Y_3\rangle$.  The forced projectors are indexed 13, 17, and 21.
To be very explicit, the rank-2 projector indexed 1, as expressed in the $Z$ eigenbasis, $\{|0\rangle=|+Z\rangle, |1\rangle=|-Z\rangle\}$, is given explicitly as,
\begin{equation}
|+A, +B, +C| = |+ZIZ, +ZZI, +IZZ|
\end{equation}
\begin{equation}
 = |+ZIZ, +ZZI, +IZZ, +ZII,\ldots| + |+ZIZ, +ZZI, +IZZ,-ZII,\ldots|
\end{equation}
\begin{equation}
= |000\rangle\langle000| + |111\rangle\langle111|.
\end{equation}

Fig. \ref{BasesColor} shows the process of assigning realist truth values to the 24 bases one step at a time, with green projectors assigned the value 1, and blue projectors assigned the value 0.  This is analogous to the information in Fig. \ref{KS}, but we present it in this second form for enhanced clarity.  To begin, since projectors 5 and 9 are the PPS, they are automatically assigned truth-value 1 ($v_e(5) = v_e(9) = 1$), and therefore projectors orthogonal to 5 and 9 are assigned truth-value 0, as shown in blue in Fig. \ref{S1}.  Next, because rows 4, 5 and 6 in Figs. \ref{KS} and \ref{S1} now contain three 0 (blue) truth-values, the ABL rule tells us that we must assign a truth-value of 1 to projectors 13,17, and 21 ($v_e(13) = v_e(17) = v_e(21) = 1$, as shown in Figs. \ref{KS} and \ref{S2}, where the underlines indicate forced values.)  Now we must assign a truth-value 0 to all projectors orthogonal to projectors 13, 17 and 21, namely projectors 1, 2, 3, and 4 ($v_e(1) = v_e(2) = v_e(3) =v_e(4) = 0$, as shown in Figs. \ref{KS} and \ref{S3}, where again underlines indicate forced values).  We can now explicitly see the violation of the sum rule (bold face rows in Figs. \ref{KS} and \ref{S3}) and this is the meaning of KS contextuality in the projector picture.

This exercise of constructing Fig. \ref{KS} shows us that if we want to assume that a complete 1 or 0 truth-value assignment exists and is consistent with both the pre- and post-selections, then we can now explicitly identify the conflict basis.  The original motivation for assigning these noncontextual truth-values was that they should predict with certainty the outcome (eigenvalue) of a measurement in any randomly chosen context.  In our model, $v_e$ depends explicitly on the PPS, and thus it only needs to predict eigenvalues for contexts that contain those two states - since no other eigenvalues will be measured.  Furthermore, aside from the extension due to the ABL rule, it is impossible to learn the realist truth-values $v_e$ of projectors in contexts that do not contain the PPS states, and thus there is no physical grounds for requiring that eigenvalues should be assigned to them at all - and as we have seen, such eigenvalue assignments $v_e$ can violate the sum rule.

On the other hand, the weak values $v_w$ perfectly match all of the eigenvalues that we learn from the PPS and ABL rule, while also obeying the sum rule in all other contexts (because they can take anomalous values).  More importantly, we {\it can} learn the weak values of projectors in contexts that do not contain the PPS states through the use of weak measurements.  This shows that the complete weak value assignment is actually much more consistent with the measurable physics of this situation than the eigenvalue (truth-value) assignment.

To illustrate this point, consider that the eigenvalues $(-,-,-)$ for the ID $(ZIZ, ZZI, IZZ)$ in the top row of Fig \ref{SquareS3} do not even correspond to an eigenstate in Hilbert space, and so there is no way within quantum theory to measure this `state' (i.e. it is impossible to define measurement operators and a POVM that contain these outcomes)  On the other hand, the maximum conflict projector $(+,+,+)$ {\it is} an eigenstate of this ID, and thus we can weakly measure its {\it anomalous} weak value, which is just the signature that provides concrete evidence of contextuality \cite{pusey2014anomalous}.

\subsubsection{Discussion}

For this example, we see that the classical basis is a conflict basis (top row of Fig. \ref{KS}) and that it contains an anomalous weak value, but not all conflict bases contain anomalous weak values, nor are conflict bases the only bases that contain anomalous weak values.  We show in the Appendices \ref{PauliRules}, \ref{KS_NQPauli}, \ref{Wheel3Derivation}, \ref{Wheel4Derivation}, and \ref{ArchDerivation} that for a large class of KS sets within the $N$-qubit Pauli group, the general condition is that if a conflict basis has all 0 realist truth-values, then no projector in that basis can have a weak value greater than 1/2, and if it is also an eigenbasis, then it must contain at least one projector with a negative weak value.  In these cases, all of the projectors in a conflict basis that do not have weak values of 0 must have real weak values of $\pm w$.  The ABL formula gives 0 probability for a projector with a weak value of 0, but it gives all of these other projectors an equal probability $1/\delta$ to be found by a strong measurement during the interval between the pre- and post-selection, where $\delta$ is the number of nonzero weak values in the measurement basis.  These equal ABL probabilities seem to be a unique feature of value-indefiniteness in KS sets, which possess enough symmetry to be state-independent proofs of contextuality.  In Appendix \ref{NonKS} we consider an alternate demonstration of the product-state pigeonhole effect for $N$-qubits that does not depend on a KS set, and find that even though the conflict basis does contain projectors with anomalous weak values, they do not all have equal magnitudes, and thus these projectors do not have equal ABL probability.

We should note that although all 24 projectors were assigned realist truth-values in this example, this does not happen for a general KS set and PPS.  There are also many choices of KS sets, and PPSs from within them, that do not force any value assignments, or even if they do, they do not force any conflict bases.  See Appendix \ref{GHZStar} for a detailed example.

In this example, the pre-selection, post-selection, and conflict, all occur in product bases, and the only role played by entanglement is in the forced bases.  In fact, the classical basis reveals an anomalous weak value even if we never make the entangled weak measurements at all, and thus the relative spatial distance between the qubit systems does not seem to matter, which may have remarkable physical implications, as discussed in \cite{aharonov2014quantum}.

It is also important to note that with a product PPS, the weak value of a classical projector can be factored into the direct product of the weak values for each individual qubit, and thus the measurements needed to find the anomalous weak value can be performed on each qubit individually - at arbitrary locations.  It may seem puzzling that this can happen even without entanglement, but to understand it, consider the single-qubit weak values that go into the anomalous projector in our example,
\begin{equation}
q_w = \frac{\langle+Y|+Z\rangle \langle+Z| +X\rangle}{\langle+Y|+X\rangle} = \frac{1}{\sqrt{2}}e^{ \frac{i\pi}{4}}.
\end{equation}
Clearly the product of three such weak values has a negative real part (and the imaginary parts cancel in the weak value of the rank-2 projector), and in this way quantum contextuality can be revealed even for product eigenstates of Pauli observables.  Outside the Pauli group there are choices of single-qubit projectors and PPSs that have negative weak values individually, and entanglement does not seem to play any role at all.

This product-state pigeonhole effect can be extended to any number of qubits using what is known as the Wheel family of KS sets, of which Fig. \ref{Square} is the simplest member \cite{WA_Nqubits, WaegellThesis, W_Primitive}.  The Wheels for odd $N$ give a direct demonstration of the pigeonhole effect (see Fig. \ref{Wheel5}, while for even $N$, we can use $N$ different overlapping $(N-1)$-qubit Wheels to show it (see Appendices \ref{Wheel3Derivation} and \ref{Wheel4Derivation}).
\begin{figure}[h!]
\centering
\caption{The 5-qubit Wheel}
\includegraphics[width=3in]{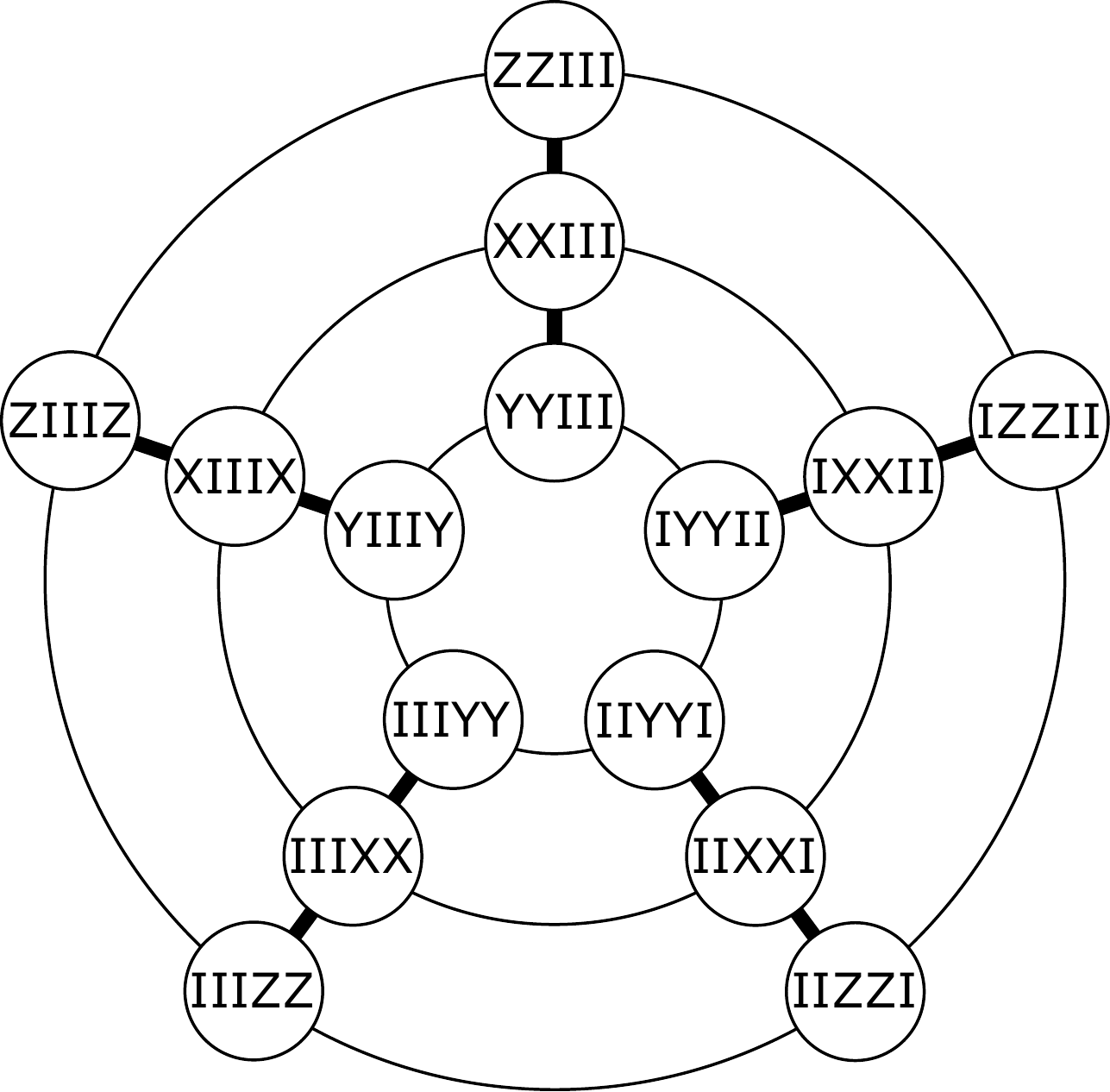}
\label{Wheel5}
\end{figure}

\begin{figure}[h!]
\centering
\caption{A 3-qubit Kite}
\includegraphics[width=3in]{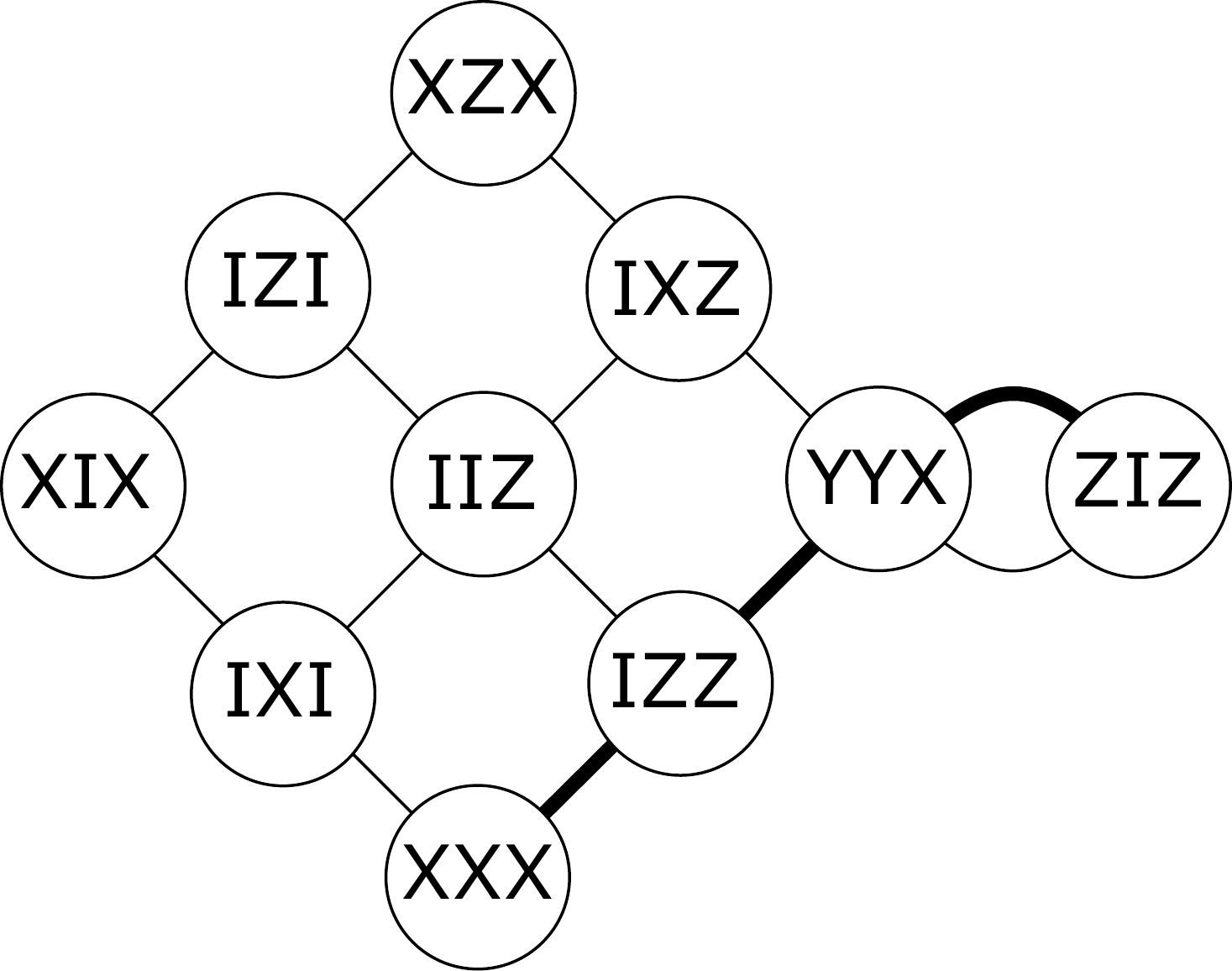}
\label{Kite3}
\end{figure}

There are many other examples wherein either the PPS or the conflict basis are entangled, including the 2-qubit Peres-Mermin Square - for which the pigeonhole effect has already been named the `Quantum Cheshire Cat,' and many others, as shown in Figs.\ref{Kite3}, \ref{Kite}, and \ref{Arch}. Choosing the right PPS for any of these KS sets, and local unitaries to permute $\{Z,X,Y\}$, one can always arrange for the conflict to occur in the classical basis.

The Arch \cite{WA_Nqubits, WaegellThesis} (Appendix \ref{ArchDerivation}) is structurally interesting in that part of the classical basis can be assigned 0 by the PPS.  The Kites are interesting because the value forced by the ABL rule can occur in an ID of any length, and furthermore the Kites are the most general family of KS sets, with Kites occurring for all distinct types of stabilizer states within the $N$-qubit Pauli group \cite{WA_Nqubits, WaegellThesis, W_Primitive, W_Bonding}.  There are many other types of KS sets within the $N$-qubit Pauli group with some or all of these characteristics.

From the examples given here, it is clear that many KS sets can be directly applied to demonstrate the $N$-qubit quantum pigeonhole effect, but there are also KS sets that cannot be used in this way, because there is no PPS that forces a conflict basis to occur in that set.  However, in order to determine if there is a KS-pigeonhole effect in the $N$-qubit Pauli group for a given PPS, we must consider all KS sets that contain IDs from within both the pre-selected and post-selected stabilizer groups, and so the failure of one particular KS set actually tells us very little.  This issue is explored in Appendix \ref{GHZStar}.

Let us now return to the ingenious value-indefiniteness arguments presented by Abbott et al \cite{Abbott2015}, and examine how they are related to the conflict bases and projectors we have been working with.  There is an obvious similarity, in that both models begin by choosing a particular projector from within some set of projectors and contexts to be assigned truth-value 1, but this is largely where the similarity ends.

In their model, they choose a second projector in the set that is neither parallel nor orthogonal to the first, and then show that assigning a truth-value 1 to that projector logically forces a violation of the sum rule, where the forcing is akin to iterative applications of the ABL rule (note that the true ABL rule cannot be iterated in this way).  Likewise, if a truth-value 0 is assigned to the this projector, this also forces a violation of the sum rule.  Thus, given the prepared state, the authors then conclude that the second projector itself cannot have a definite value that belongs to complete definite truth-value assignment to all projectors in the set, and that this projector must therefore be value-indefinite.  Again, the complete definite truth-value assignments is only required because of the assumption that we can make a measurement in randomly chosen context, that could therefore reveal any portion of the assignment.  The authors also pay no special attention to the particular basis where the sum rule violation is forced, nor should they, since for a general set of projectors the answer could be different for different choices of the second projector, or even arbitrary different orders of applying their forcing rule to obtain the violation.  Under the assumption of random choice, they are correct to conclude that any legitimate choice of a second projector from the set must be value-indefinite.

In contrast, by including a post-selection in addition to the pre-selection, our model explicitly assigns truth-value 1 to a second projector in the set.  Then, using only the ABL rule and our realist truth-value assignment, with the assumption of PPS-contextuality (which can in some sense be seen as complementary to random choice), we are able to force a violation of the sum rule.  In our model, only the projectors in a conflict basis are explicitly value-indefinite, and the weak values of projectors in such a basis reveal a clear signature of this indefiniteness.  Again, we should stress that in our model there is no fully random choice, or if there is, then the entire truth-value assignment to the set responds to it retrocausally.

\subsection{Conclusion}

We have given a straightforward interpretation of KS contextuality within the framework of the ABL reformulation of quantum mechanics, and introduced the notion of a realist truth-value assignment to the observables of a pre- and post-selected quantum state.  From there, we showed how weak values extend the amount of information we can learn about a given PPS state, and how this can be used to reveal more of the realist truth-value assignment, and to identify conflict bases.  We have also shown how the weak values of the projectors in a conflict basis of a KS set can experimentally reveal the signatures of the conflict, which are anomalous weak values and ABL value-indefiniteness.

This analysis also reveals an important feature of the ABL reformulation with respect to contextuality.  In the ABL reformulation of quantum mechanics, we presume the existence of a pre-selected state and a post-selected state that are both legitimate quantum states - which is also to say that the eigenvalues and observables that define these states cannot violate the sum or product rule.  This means that the ABL reformulation already implicitly requires that if there is a context where the sum or product rule is violated, it is never a context that is fully defined by the pre- or post-selected state - which is just the hidden variable model ($v_e$) we have been exploring, complete with the free-choice/retrocausality condition introduced above.

We have shown how the quantum pigeonhole principle for $N$-qubits is demonstrated by structural proofs of contextuality within the $N$-qubit Pauli group.  Furthermore, because the pigeonhole effect based on the Wheel family has product states for the PPS and conflict basis, we can experimentally demonstrate the quantum pigeonhole effect for $N$ qubits by measuring the weak values in the classical basis, regardless of how far apart the qubits are located when the measurements are performed.  The only place that entanglement appears in this scenario is that we would need weak measurements in entangled bases in order to explicitly verify the ABL rule.  Whether or not we make those entangled weak measurements, the weak value signature of the conflict bases is always obtained by making local measurements.

We should mention that a similar 2-qubit PPS paradox has been explored by Cabello \cite{cabello1997no} and shown to be inconsistent with noncontextual hidden variable models - even without making use of any complete KS set.  This example differs from the pigeonhole effect in that the conflict does not occur in a classical basis, but rather in a maximally entangled basis.  Hardy \cite{hardy1992quantum} gave a similar proof using a maximally entangled pre-selection state, and in this case, the conflict occurs in a product basis, which can be converted into the classical basis by applying suitably chosen local unitary operations to the entire set.  Furthermore, while this example uses the ABL rule to force certain assignments (though the author refers only to the sum rule), no use is made of the weak values of the projectors in the set.  It is trivial to see from the sum rule that any conflict basis with two or more projectors with forced realist truth-values of 1 necessarily contains another projector with a negative weak value, and so Pusey's new signature of contextuality \cite{pusey2014anomalous} is also present in these examples.

\textbf{Acknowledgements:}  We thank P.K. Aravind, Sandu Popescu, and Yakir Aharonov, for several helpful discussions.  Funding for this research was provided by the Institute for Quantum Studies at Chapman University and the Fetzer Franklin Fund.

\bibliographystyle{ieeetr}
\bibliography{Rabbithole.bbl}
\subsection{Appendix}

In the following appendices we derive some general formulae for working with the ABL reformulation, weak values, stabilizer states, IDs, and KS sets within the $N$-qubit Pauli group, and how to use them to show the pigeonhole effect.  Using these formulae, we proceed to carefully work through several important examples, deriving other results along the way.

\subsubsection{The ABL Reformulation of Quantum Mechanics} \label{ABLDerivation}

 Here we give a brief review and generalization of some relevant parts of the ABL formalism.

 To begin, let us define some general notation to refer to projectors of different rank.  We define a rank-1 projector as
 \begin{equation}
 \Pi_i^1 \equiv |i\rangle\langle i|,
 \end{equation}
 and subsequently define a rank-$r$ projector in terms of rank-1 projectors as
 \begin{equation}
 \Pi_i^{r_i} = \sum_j^{r_i} |j_i\rangle\langle j_i| = \sum_j^{r_i} \Pi_{ij}^1.
 \end{equation}
 where $\{|j_i\rangle\}$ is any orthonormal basis (kets forming rank-1 projectors) that spans that $r_i$-dimensional subspace $\Pi_i^{r_i}$ projects onto.  The arbitrary choice of $\{|j_i\rangle\}$ is analogous to preparation-independence of mixed-state density matrices.

 We define a complete measurement basis $\vec{B}$ on a system of Hilbert-space dimension $d$, as any set of mutually orthogonal projectors that span the space - regardless of their individual ranks.
 \begin{equation}
 \vec{B} = \{\Pi_1^{r_1},\Pi_2^{r_2},\ldots,\Pi_{d'}^{r_{d'}}\},
 \end{equation}
 where $d'$ is the cardinality of the set $\vec{B}$, $\sum_i^{d'} r_{i} = d$, and $\sum_i^{d'} \Pi_{i}^{r_i} = I$.  Thus it is clear that $\vec{B}$ completely defines the POVM that will actually be measured during an experiment.  This experiment truly measures all observables that have $\vec{B}$ as an eigenbasis, which is obvious when such an observable is spectrally decomposed,
\begin{equation}
A = \sum_i^{d'} a_i \Pi_{i}^{r_i},
\end{equation}
where $a_i$ are the eigenvalues.

For pre-selected state $|\Psi\rangle$, post-selected state $|\Phi\rangle$, the weak value of a rank-1 projector is given by,
\begin{equation}
(\Pi_i^1)_w = \frac{\langle \Phi | i \rangle \langle i | \Psi\rangle}{\langle \Phi|\Psi\rangle},
\end{equation}
whereas for a rank-$r$ projector it is,
\begin{equation}
(\Pi_i^{r_i})_w = \frac{\sum_j^{r_i}  \langle \Phi | j_i \rangle \langle j_i | \Psi\rangle}{\langle \Phi|\Psi\rangle} = \sum_j^{r_i} (\Pi_{ij}^1)_w.
\end{equation}
Note that weak values of projectors are defined without reference to any specific basis $\vec{B}$, which is one common definition of noncontextuality.  It is important to note that the weak values also obey the sum rule, $\sum_i^{d'} (\Pi_{i}^{r_i})_w = 1$, for all possible bases $\vec{B}$.

Now, the ABL formula gives the probability to obtain a particular outcome when an intermediate measurement is made in basis $\vec{B}$ between the pre-selection of $|\Psi\rangle$ and post-selection of $|\Phi\rangle$.
\begin{equation}
P(|\Pi_i^{r_i}=1 \space  / |\Psi\rangle, |\Phi\rangle, \vec{B}) = \frac{  |\sum_j^{r_i} \langle \Phi | \Pi^1_{ij} | \Psi\rangle|^2  } { \sum_k^{d'} | \sum_j^{r_k}\langle \Phi | \Pi^1_{kj} | \Psi\rangle|^2 },
\end{equation}
\begin{equation}
 = \frac{  |\sum_j^{r_i} (\Pi^1_{ij})_w|^2  } { \sum_k^{d'} | \sum_j^{r_k} (\Pi^1_{kj})_w|^2 } = \frac{  |(\Pi^{r_i}_{i})_w|^2  } { \sum_k^{d'} | (\Pi^{r_k}_{k})_w|^2 },\label{ABLFormula}
\end{equation}

One of the reasons we have emphasized the ranks of projectors in this discussion is that the ABL probabilities cannot be coarse-grained by summation as can Born rule probabilities.  Consider the case, $\Pi^1_1 + \Pi^1_2 = \Pi^2$.  Because they span the same space, either the two projectors on the left, or the one on the right, but not both, can be contained in a basis $\vec{B}$.  In general, the denominator of Eq. \ref{ABLFormula} can be different for the two alternative bases, and thus one cannot conclude that $P(|\Pi_1^1=1 \space  / |\Psi\rangle, |\Phi\rangle, \vec{B}) + P(|\Pi_2^1=1 \space  / |\Psi\rangle, |\Phi\rangle, \vec{B}) = P(|\Pi^2=1 \space  / |\Psi\rangle, |\Phi\rangle, \vec{B})$.

Now we have everything we need to derive the ABL rule.  If the ABL formula gives probability 1 to obtain $\Pi_i^{r_i} = 1$ when measured in basis $\vec{B}$, it follows that $\langle \Phi | \Pi_k^{r_k} | \Psi\rangle = 0$ for all $k \neq i$.  Then from the sum rule for weak values we find that,
\begin{equation}
(\Pi_i^{r_i})_w = \frac{\langle\Phi | \Pi_i^{r_i} | \Psi\rangle}{\langle \Phi|\Psi\rangle} = 1,
\end{equation}
which is the ABL rule.  Specifically, if the ABL formula predicts unit probability to obtain a given projector by an intermediate strong measurement, then that projector has a weak value of 1.  It is also trivial to see that when the ABL formula predicts probability 0 for a given projector, then that projector has a weak value of 0.

The reverse ABL rule is slightly more subtle.  Given that $(\Pi_i^{r_i})_w=1$, then from the sum rule $\sum_{k \neq i}^{d'} (\Pi_k^{r_k})_w = 0$, for any basis that contains $\Pi_i^{r_i}$.  If $d'=2$ for some basis $\vec{B}$, then we know both weak values, and we can see that the ABL formula gives unit probability to obtain outcome $\Pi_i^{r_i}=1$, which is the reverse ABL rule.  For larger cardinalities $d' = 2+n$, it must also be given that $(\Pi_k^{r_k})_w=0$ for $n$ of the projectors in $\vec{B}$ in order to ensure unit probability to obtain $\Pi_i^{r_i}=1$.  It is also trivial to see that if the weak value of a projector is 0, then the ABL formula gives 0 probability to obtain that outcome by an intermediate measurement, and this is true for systems of all dimensions.  It is important to stress that the applicability of the reverse ABL rules depends on the cardinality of $\vec{B}$, and not on the spectrum of the observables in question.

The rules for projectors can be applied to obtain rules for IDs.  In this case, if all but one observable in an ID has definite eigenvalues from the pre and/or post-selected state, then the ABL rule forces the last observable to have a value such that the product matches the sign of the ID.  Likewise, because all $N$-qubit Pauli observables are dichotomic, then if the weak value of such an observable is an eigenvalue, then by the reverse ABL rule, the outcome of an intermediate measurement that determines only the eigenvalue of that single Pauli observable (and no others, i.e. a parity measurement with $d'=2$) will always agree with the weak value.

A final issue of the ABL formulation has been referred to elsewhere as the situation of {\it diagonal PPS} \cite{tollaksen2007pre}, which relates to sequential measurements of mutually commuting observables during the interval between pre- and post-selection.  Specifically the issue arises when several mutually commuting observables have values forced by the ABL rule.  An intermediate measurement of any one of these observables will yield the forced value with certainty, but if more than one of them is measured sequentially, the extra measurements break the interval into several concatenated subintervals, disturbing the original PPS, and the original ABL rule no longer applies.  This disturbance occurs even despite the fact that the sequentially measured observables mutually commute.

We will use the 3-qubit square and pigeonhole effect described above to illustrate why this happens.  Given the $|\Psi\rangle$ and $|\Phi\rangle$, we have $(XXI)_w = 1$, and $(YYI)_w = 1$, and thus the ABL rule forces $(ZZI)_w = -1$ because this vertical ID is negative.  A intermediate parity measurement ($d'=2$) of $ZZI$ must then yield -1 as an outcome given this PPS.  The same is true for $ZIZ$ and $IZZ$.  If however, we were to measure both $ZZI$ and $ZIZ$ sequentially during the intermediate time, because the two parity measurements commute, the combined effect is to measure a basis with $d'=4$ which projects the state onto an eigenstate of the classical ID $(ZZI, ZIZ, IZZ)$, and the pigeonhole effect is lost.  On the other hand, we can perform sequential parity measurements of the observables in a the Bell ID $(XXI, ZZI, YYI)$ that forces the value of $ZZI$.  If we sequentially measure $XXI$, $ZZI$, and $YYI$, we will project the first two qubits onto a Bell state, and the pigeonhole effect remains.  Diagonal-PPS refers to the fact that one ID can be sequentially measured during the intermediate time without disturbing the PPS while the other cannot, and depends on how the stabilizer groups of the PPS commute with the sequence of observables.

\subsubsection{The Mean King's Problem} \label{MeanKing}

The examination of the weak values of all projectors in a KS set is also useful for seeing why the Mean King \cite{vaidman1987ascertain, englert2001mean, aravind2003solution} will always win the game if he chooses measurements from a KS set.  In order for the physicist (the other player of the game) to succeed, he must be able to predict with certainty the King's outcome for an intermediate measurement of any observable from among a specified set of observables - that need not mutually commute.  In other words, he must be able to choose a special pre-selected state and post-selected basis such that for each measurement outcome the King can obtain, the ABL formula predicts unit probability for that outcome, and zero for all other other outcomes in that measurement basis.  From the ABL rule, it then follows that every projector the suitor can measure must have a weak value of 0 or 1.  Now, a KS set by definition cannot admit {\it any} noncontextual truth-value assignment of 0s and 1s to all projectors in the set without the sum rule being violated in some bases.  As we have discussed above, the weak values of the projectors in any set, $v_w$, are noncontextual and obey the sum rule by definition, and therefore they cannot all be 0 and 1 for any KS set.  It then follows from Eq. \ref{ABLFormula} that if the Mean King is allowed to make measurements from a KS set then he will always win the game, since there can be no pre- and post-selection for which the ABL probabilities to obtain the projectors by an intermediate measurement are all 0 or 1.

This simple new argument using weak values generalizes Mermin's original observation \cite{mermin1995limits} to all possible KS sets, and indeed to any set for which all 0/1 weak values are impossible.  This fails to include many state-dependent proofs of contextuality, but it does apply to the 13-ray KS proof of Yu and Oh \cite{yu2012state}.  This is an interesting case because the set is colorable in the KS sense, but there is a geometric feature common to all legitimate colorings of the set that disagrees with a quantum mechanical prediction, allowing the KS theorem to be proved.  This suggests that there could be a novel method for using weak values to identify many new colorable sets that prove the KS theorem, which will be worth exploring in the future.

As a corollary, The Yu-Oh set is based on a SIC-POVM, and this idea has been generalized \cite{bengtsson2012kochen}.  Generalizing in a different direction, we can see that whenever the elements of the POVM can be scaled to be projectors that sum to a non-integer multiple of identity, it is impossible to choose a PPS such that the weak values of all of these projectors are 0 or 1.

Finally we present an open question that follows from these ideas and attempts to go back the other way.  Let $\{\Pi_i\}$ be any set of projectors that satisfies the condition that it is possible to assign noncontextual values 0 or 1 to all of these projectors such that the sum rule is obeyed for all scaled POVMs formed by $\{\Pi_i\}$ (i.e. the most general non-KS set).

Can it be shown that for any such set, there exists a PPS such that $\{(\Pi_i)_w\} = \{0,1\}$ for all $i$ in the set?  If the answer is positive, it follows that any set of projectors for which no PPS can give $\{(\Pi_i)_w\} = \{0,1\}$ is a proof of the Kochen-Specker theorem - including the non-integer POVMs discussed above.  This would be a remarkable result, and would widely expand the domain of contextual quantum mechanics.

It is worth mentioning that this application of the weak value has absolutely nothing to do with weak measurements - the weak value is simply a mathematical property of the pre- and post-selected quantum system that helps to make the situation more transparent.


\subsubsection{The $N$-qubit Pauli Group} \label{PauliRules}

The first rule to note is that any two $N$-qubit Pauli observables must either commute or anticommute.  As a result, if one observable is applied as a similarity transformation to another, then if the two observables commute the transformation does nothing, and if they anticommute it introduces a negative sign.

Let us now consider what happens if we act with an $N$-qubit Pauli operation $U$ on such an $N$-qubit Pauli state.  Applying a unitary to a quantum state also applies that unitary as a similarity transformation to all of the observables in that state's stabilizer group.  We must also be careful to track the relative phase $a$ that is introduced by such a transformation.
\begin{equation}
U|\lambda_A A, \lambda_B B, \lambda_C C \rangle
\end{equation}
\begin{equation}
= a|\lambda_A a U A U a^*, \lambda_B a U B U a^*, \lambda_C a U C U a^* \rangle
\end{equation}
\begin{equation}
=a|c_{AU}\lambda_A A, c_{BU}\lambda_B B, c_{CU}\lambda_C C \rangle = a|\lambda_A' A, \lambda_B' B, \lambda_C' C \rangle
\end{equation}
where $c_{QP} = 1$ if $[Q,P]=0$ and $c_{QP} = -1$ if $\{Q,P\}=0$, and these signs are simply absorbed into the eigenvalues of the new state as $\lambda_i' = c_{iU}\lambda_i$.  Thus for all possible $a U$, the state remains an eigenstate of the same stabilizer group - though it may now be a different state in the eigenbasis.  For $U=A$, it is easy to see that $a=\lambda_A$.

At this point we can derive a useful formula that we will need later.  Suppose that $V$ is another Pauli observable, such that $c_{QU}=c_{QV}$ for all $Q=\{A,B,C\}$, so that,
\begin{equation}
V|\lambda_A A, \lambda_B B, \lambda_C C \rangle = b|\lambda_A' A, \lambda_B' B, \lambda_C' C \rangle.
\end{equation}
Furthermore suppose that $(A,U,V)$ is an ID with sign $s_{_{AUV}}$.  Taking the inner product of the last two expressions we obtain,
\begin{equation}
ab^* = \langle \lambda_A A, \lambda_B B, \lambda_C C|VU|\lambda_A A, \lambda_B B, \lambda_C C \rangle
\end{equation}
\begin{equation}
= s_{_{AUV}}\langle \lambda_A A, \lambda_B B, \lambda_C C|A|\lambda_A A, \lambda_B B, \lambda_C C \rangle
\end{equation}
\begin{equation}
= \lambda_A s_{_{AUV}}.  \label{UV}
\end{equation}

We can use this formula to find relations between the weak values of different projectors in the same eigenbasis, and in some cases this is enough to calculate all of the weak values exactly.

\subsubsection{KS sets within the $N$-qubit Pauli group} \label{KS_NQPauli}

There exists a family of KS sets composed of observables and contexts from within the $N$-qubit Pauli group.  The observables are simply tensor products of the Pauli spin matrices and the identity, while the contexts are sets of mutually commuting $N$-qubit observables.  A maximal set of mutually commuting observables is called a stabilizer group, and we will further call any set of mutually commuting observables that is closed under multiplication a sub-stabilizer group.  Within a given stabilizer group, one can find minimal sets of $M$ observables whose product is $\pm I^{\otimes N}$ (identity in the space of all $N$ qubits).  Such sets contain $M-1$ independent generators, and the last observable is the product of those.  We call these minimal subgroups IDs or {\it Identity Products}, and characterize them by the symbol ID$M^N$ for a set of $M$ observables from the $N$-qubit Pauli group \cite{WaegellThesis, W_Primitive}.  Any ID is also a measurement context.

A minimal KS set \cite{WA_3qubits, WA_4qubits, WA_Nqubits}, is a set of observables and IDs with the following properties: 1) each observable in the set appears in an even number of the IDs in the set, and 2) the number of IDs in the set whose product is $-I$ (negative IDs) is odd.

To see how such a set proves the KS theorem, consider the overall product of all of the noncontextual truth-values ($\pm 1$) for all the IDs in the set.  If the product rule is to be obeyed, then this product must be -1, since there are an odd number of negative IDs in the set.  However, the truth-value for each observable appears an even number of times in this product, since each observable belongs to an even number of IDs in the set, and thus noncontextuality implies that the product must be 1, and the theorem is proved.

\begin{figure}[h!]
\centering
\caption{A 5-qubit Kite}
\includegraphics[width=3in]{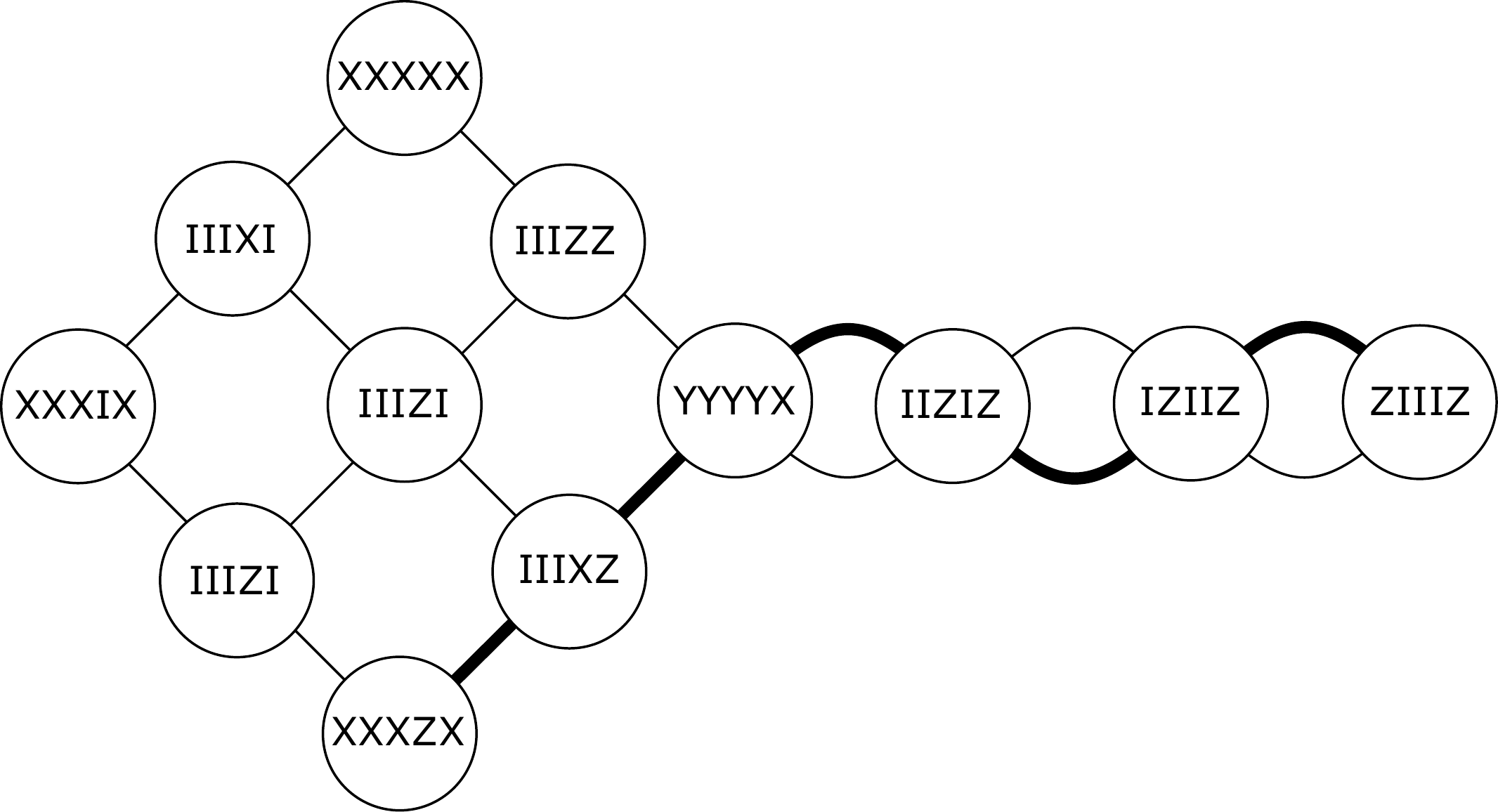}
\label{Kite}
\end{figure}

There are many examples of such KS sets based on observables from the $N$-qubit Pauli group.  In Figs. \ref{Square}, \ref{Wheel5}, \ref{Arch}, \ref{Kite3}, and \ref{Kite}, we give diagrams for some example KS sets that can show the pigeonhole effect for certain pre-and post-selections, as discussed below.  The observables along a line or arc form an ID, and thick lines denote negative IDs, while thin lines denote positive IDs.  These diagrams make the proof of the KS theorem transparent because it is easy to see that each observable lies in an even number of lines, while the number of thick lines is odd.

We can obtain another type of KS set composed of projectors, and contexts that are complete orthogonal bases of these projectors.  A KS set of projectors and bases is a set that cannot admit a noncontextual truth-value assignment (0 or 1) to the projectors in the set in such a way that the sum rule is satisfied for all of the bases.  Furthermore, a {\it parity KS set} is a set of projectors and bases in which each projector in the set appears in an even number of the bases in the set, and the total number of bases is odd \cite{WA_24Rays, WA_60Rays, Cabello18_9, lisonvek2014kochen, WA_600cell_2, MP_600cell, WA_120cell, lisonek2014generalized}.

To see how a parity KS set proves the KS theorem, consider the overall sum of all of the noncontextual truth-values for all of the bases in the set.  If the sum rule is to be obeyed, this sum must be equal to the odd number of bases.  However, the truth-value for each observable appears an even number of times in this sum, and thus noncontextuality implies that the sum must be even, and the theorem is proved.

A projector-based KS set is also called {\it critical}, or minimal, if the KS proof would fail if any context were removed from the set.

Each ID also gives rise to a joint eigenbasis of projectors.  The complete set of projectors taken from all IDs in a KS set of the former (observables) type always gives a KS set of the latter (projectors) type.  This KS set is never a critical parity KS set, but always contains them as subsets.

\subsubsection{3-qubit Square (or Wheel)} \label{Wheel3Derivation}
Let us now return to, and generalize, the example case of the 3-qubit square KS set.  Specifically we are interested in the weak values of the projectors in the conflict basis.  Our pre- and post-selected states are
\begin{equation}
|\Psi\rangle \equiv |\lambda_D D, \lambda_E E, \lambda_F F\rangle,
\end{equation}
and
\begin{equation}
|\Phi\rangle = |\lambda_G G, \lambda_H H, \lambda_I I\rangle,
\end{equation}
such that $s_{_{DEF}} = \lambda_D  \lambda_E \lambda_F = +1$ and $s_{_{GHI}}=\lambda_G \lambda_H \lambda_I = +1$.  This PPS always leads to a conflict in the classical basis.  We define the maximum conflict projector for which all eigenvalues are opposite to the forced values.
\begin{equation}
|\mathcal{C}| = |\lambda_D \lambda_G A, \lambda_E \lambda_H  B, \lambda_F \lambda_I C|,
\end{equation}
\begin{equation}
\equiv |\lambda_A A, \lambda_B B, \lambda_C C|,
\end{equation}
noting also that $s_{_{ABC}} = \lambda_A \lambda_B \lambda_C = +1$ (which can be verified by direct substitution of $\lambda_A = \lambda_D \lambda_G,\ldots$).

The weak value of this projector is
\begin{equation}
|\lambda_A A, \lambda_B B, \lambda_C C|_w = \frac{\langle \Phi | \mathcal{C} | \Psi \rangle}{\langle \Phi | \Psi \rangle}
\end{equation}
\begin{equation}
 = \frac{\langle\lambda_G G, \lambda_H H, \lambda_I I|\lambda_A A, \lambda_B B, \lambda_C C|\lambda_D D, \lambda_E E, \lambda_F F\rangle}{\langle\lambda_G G, \lambda_H H, \lambda_I I|\lambda_D D, \lambda_E E, \lambda_F F\rangle}.
\end{equation}

Expanding $|\mathcal{C}|$ as a sum of rank-1 projectors and using Eq. \ref{UV}, we can see that
\begin{equation}
|\lambda_A A, \lambda_B B, \lambda_C C| = \lambda_A s_{_{ADG}} (G|\lambda_A A, -\lambda_B B, -\lambda_C C|D)
\end{equation}
\begin{equation}
=\lambda_B s_{_{BEH}} (H|-\lambda_A A, \lambda_B B, -\lambda_C C|E)
\end{equation}
\begin{equation}
=\lambda_C s_{_{CFI}} (I|-\lambda_A A, -\lambda_B B, \lambda_C C|F),
\end{equation}
where we have made explicit use of the commutation relations among these nine observables in order to obtain the values of $c_{QP}$.

Taking the weak value we obtain,
\begin{equation}
|\lambda_A A, \lambda_B B, \lambda_C C|_w = \lambda_A \lambda_D \lambda_G s_{_{ADG}}(|\lambda_A A, -\lambda_B B, -\lambda_C C|_w)
\end{equation}
\begin{equation}
=\lambda_B \lambda_E \lambda_H  s_{_{BEH}} (|-\lambda_A A, \lambda_B B, -\lambda_C C|_w)
\end{equation}
\begin{equation}
=\lambda_C \lambda_F \lambda_I  s_{_{CFI}} (|-\lambda_A A, -\lambda_B B, \lambda_C C|_w),
\end{equation}
a relation between the weak values of all four projectors in the conflict eigenbasis of the ID $(A,B,C)$, and note that $\lambda_A \lambda_D \lambda_G s_{_{ADG}}=\lambda_B \lambda_E \lambda_H  s_{_{BEH}} = \lambda_C \lambda_F \lambda_I  s_{_{CFI}} -1.$  We see that all four weak values have the same magnitude and differ only up to a sign.  From this, and the fact that these four weak values obey the sum rule (i.e. they add up to 1), we find that all four weak values are real and have magnitude 1/2, and that the maximum conflict projector $|\mathcal{C}|$ has a negative (anomalous) weak value.

Finally, noting that the product of the signs of all four weak values is,
\begin{equation}
\lambda_A \lambda_D \lambda_G\lambda_B \lambda_E \lambda_H\lambda_C \lambda_F \lambda_I  s_{_{ADG}} s_{_{BEH}} s_{_{CFI}}
\end{equation}
\begin{equation}
=s_{_{ABC}}s_{_{DEF}}s_{_{GHI}} s_{_{ADG}} s_{_{BEH}} s_{_{CFI}} = -1,
\end{equation}
we see that the negative weak value occurs exactly because this is a KS set, and the product of the signs of all IDs in a KS set must be -1.\newline\newline

It is important to point out that we obtained this result, even up to the exact weak values, without making use of the Hilbert-space form of the Pauli operators.  Only the Pauli algebra and the structure of the KS set itself is required to obtain this result.  Furthermore, the anomalous weak value follows directly from the fact that the KS set must have an odd number of negative IDs.  To understand why this is important, consider that the proof we have just given applies just as well to the 2-qubit Peres-Mermin Square as it does to the 3-qubit version, and to any other square configuration one might find - regardless of what observables and IDs may be used to legally populate that structure.  It is also interesting to note that unlike the 3-qubit cases, the anomalous projector for the 2-qubit square depends on entanglement between the qubits.

The same method can be generalized for several other classes of KS structures to find the exact weak values.  The details vary, but there are two general features that are always obeyed within these classes.  First, in a conflict basis, the maximum real part of the weak value is 1/2, and second, every conflict eigenbasis contains at least one anomalous weak value.  In all cases, the calculation can be performed entirely in the observable-based manner we have demonstrated.

The Kite family of KS sets is by far the most general and numerous, and the 2-qubit Peres-Mermin Square is the simplest member.  Unsurprisingly, the complete logic presented above extends to this entire family of KS sets (for all $N$), with the same four weak values as for the square.  For this argument to hold with certain PPSs, the tail IDs must be reduced by multiplying all tail observables together, which yields eigenbases of higher-rank projectors that may be less entangled - essentially reducing the Kite to a Square.

We can also give the general solution for the complete family of Wheels for all $N$, of which the 3-qubit Peres-Mermin Square is the simplest member. We should point out that the Wheel set is only a KS set for odd $N$, while for even $N$ it fails to have an odd number of negative IDs.  Given the same product PPS as before, now generalized to arbitrary $N$, we find that the Wheel for even $N$ is also noncolorable because of the 3-qubit Wheels that overlap with it.  It is then no surprise that we find a classical projector with an anomalous weak value for Wheels of all $N$.

The case of Wheels with odd $N$ is a trivial generalization of the 3-qubit case given above.  In each case, the magnitude of the weak values is given by $w_N = 1/2^{\frac{N-1}{2}}$.  The weak values of all $2^{N-1}$ rank-2 projectors in the classical basis must add up to 1, and therefore there must be $2^{\frac{N-1}{2}}$ more positive weak values than negative ones.  This means that the number of negative weak values in the basis is $\nu_N = 2^{N-2} -2^{\frac{N-3}{2}}$.  It is interesting that while the magnitude of the anomalous weak values decreases exponentially with $N$, the {\it cumulative anomaly}, which we define as the sum of all negative weak values in the basis, increases with $N$ as $\mathcal{A}_N =\nu_N w_N = 2^{\frac{N-3}{2}} - 1/2$.

The case of Wheels with even $N$ is slightly more complicated.  In these cases, applying the observables of the IDs does not allow one to cycle a given eigenstate into every other eigenstate in the basis.  Instead, these operations define two distinct orbits of $2^{N-2}$ projectors, with half of the eigenbasis in each.  It can be shown that the weak values of all of the projectors in one of the orbits are zero, while the weak values in the other orbit are identical to the (odd) $N-1$ case.

\subsubsection{4-qubit Wheel} \label{Wheel4Derivation}
We demonstrate the calculation for the 4-qubit case, from which the generalization to all even $N$ is straightforward.  The 4-qubit Wheel can be represented as shown in Fig. \ref{Wheel4}.  Note that unlike the horizontal IDs of Fig. \ref{Square}, the rows of Fig. \ref{Wheel4} are not complete sub-stabilizers, and we can obtain additional observables by taking products of those within each ID.  We define a {\it generalized set} as a set in which all sub-stabilizers are complete.  Completing the sub-stabilizers of Fig. \ref{Wheel4} adds three observables to each row, forming two more negative ID columns and one positive ID column.  The generalized 4-qubit Wheel set now contains four overlapping 3-qubit Wheels as explicit subsets, and in this way the generalized set proves the KS theorem.
\begin{figure}[h!]
\centering
\subfloat[][Pauli Observables]{
\begin{tabular}{|c|c|c|c|}
\hline
$ZZII$ & $IZZI$ & $IIZZ$ & $ZIIZ$ \\
\hline
$XXII$ & $IXXI$ & $IIXX$ & $XIIX$ \\
\hline
$YYII$ & $IYYI$ & $IIYY$ & $YIIY$ \\
\hline
\end{tabular}}
\qquad
\subfloat[][Labels]{
\begin{tabular}{|c|c|c|c|}
\hline
$A$ & $B$ & $C$ & $D$ \\
\hline
$E$ & $F$ & $G$ & $H$ \\
\hline
$I$ & $J$ & $K$ & $L$ \\
\hline
\end{tabular}}
\caption{The 4-qubit Wheel set.  Each row is a positive ID4.  Each column is a negative ID3.}\label{Wheel4}
\end{figure}

The pre-selected state is $|\Psi\rangle = |\lambda_E E,\lambda_F F,\lambda_G G,\lambda_H H\rangle$ and the post-selected state is $|\Phi\rangle = |\lambda_I I,\lambda_J J,\lambda_K K,\lambda_L L\rangle$, such that $\lambda_E\lambda_F\lambda_G\lambda_H = s_{_{EFGH}}=+1$ and $\lambda_I\lambda_J\lambda_K\lambda_L = s_{_{IJKL}}=+1$.   Because the 4-qubit Wheel is not a KS set, there is no explicit conflict basis.  Instead there is a particular state in the classical basis that seems to be forced (this is a secondary forcing, not an ABL rule forcing),
\begin{equation}
|\mathcal{F}| =  |-\lambda_E\lambda_I A,-\lambda_F\lambda_J B,-\lambda_G\lambda_K C,-\lambda_H\lambda_L D|
\end{equation}
\begin{equation}
\equiv  |\lambda_A A, \lambda_B B,\lambda_C C,\lambda_D D|,
\end{equation}
with $\lambda_A\lambda_B\lambda_C\lambda_D = s_{_{ABCD}}=+1$ (which can be verified by direct substitution of $\lambda_A = -\lambda_E\lambda_I, \ldots$).
As above, we will find a relation between the weak values of this projector and several others in its basis,
\begin{equation}
|\lambda_A A, \lambda_B B,\lambda_C C,\lambda_D D|_w
\end{equation}
\begin{equation}
 = \lambda_A\lambda_E \lambda_I s_{_{AEI}}|\lambda_A A, -\lambda_B B,\lambda_C C,-\lambda_D D|_w
\end{equation}
\begin{equation}
=\lambda_B\lambda_F \lambda_J s_{_{BFJ}}|-\lambda_A A, \lambda_B B,-\lambda_C C,\lambda_D D|_w
\end{equation}
\begin{equation}
 = -\lambda_A\lambda_B\lambda_E \lambda_I\lambda_F \lambda_J s_{_{AEI}}s_{_{BFJ}}|-\lambda_A A, -\lambda_B B,-\lambda_C C,-\lambda_D D|_w,\label{Anomaly}
\end{equation}
where we have used the explicit values of $c_{QP}$ to obtain the signs of the eigenvalues.  Note also that in this case $\lambda_A\lambda_E \lambda_I s_{_{AEI}} = \lambda_B\lambda_F \lambda_J s_{_{BFJ}} = +1$.  It is important to note that Eq. \ref{Anomaly} was obtained by using the two IDs $(A,E,I)$, $(B,F,J)$ in sequence, which is identical to multiplying their observables together into a new ID $(AB,EF,IJ)$ and applying it.  These three IDs then form one of the 3-qubit Wheels within the generalized 4-qubit Wheel set.

Applying $(C,G,K)$ or $(D,H,L)$ reproduces the same orbit of four eigenstates shown above, and so we omit these cases.

What about the other four states in this basis?  Let us begin with an arbitrary element that did not belong to the first orbit,
\begin{equation}
|\mathcal{G}| =  |\lambda_A A, \lambda_B B,-\lambda_C C,-\lambda_D D|.
\end{equation}
Again we find the relation between the weak values of this projector and several others in its basis,
\begin{equation}
|\lambda_A A, \lambda_B B,-\lambda_C C,-\lambda_D D|_w
\end{equation}
\begin{equation}
= \lambda_A\lambda_E \lambda_I s_{_{AEI}}|\lambda_A A, -\lambda_B B,-\lambda_C C,\lambda_D D|_w
\end{equation}
\begin{equation}
=\lambda_B\lambda_F \lambda_J s_{_{BFJ}}|-\lambda_A A, \lambda_B B,\lambda_C C,-\lambda_D D|_w
\end{equation}
\begin{equation}
 = -\lambda_A\lambda_B\lambda_E \lambda_I\lambda_F \lambda_J s_{_{AEI}}s_{_{BFJ}}|-\lambda_A A, -\lambda_B B,\lambda_C C,\lambda_D D|_w,
\end{equation}
This is our second orbit of four eigenstates, obtained using the same method as before.

Before we use these relations to compute weak values, we need to note that some of the eigenstates in this basis are orthogonal to the PPS, but in a way that is not immediately obvious.  In particular, we note that $R \equiv AC=BD=ZZZZ$.

To construct rank-2 projectors, one takes a sum of any two orthogonal rank-1 projectors within the space of the rank-2 projector.  One way to find two orthogonal rank-1 projectors within the space of the rank-2 projectors in the classical basis is to add any new observable that commutes with, and is independent of, $(A,B,C,D)$.  We will choose to add the observable $S = XXXX$, and the rank-2 projector will take the following form,
\begin{equation}
|\lambda_A' A, \lambda_B' B,\lambda_C' C,\lambda_D' D|
\end{equation}
\begin{equation}
= |\lambda_A' A, \lambda_B' B,\lambda_C' C,\lambda_D' D, +S|+ |\lambda_A' A, \lambda_B' B,\lambda_C' C,\lambda_D' D, -S|.\label{Rank2}
\end{equation}
Noting that the sub-stabilizer already contains $R \equiv AC=BD=ZZZZ$, once we add in $S$ the expanded stabilizer now also contains $T \equiv RS = YYYY$ and the positive ID $(R,S,T)$.  Omitting all of the identical labels for $(A,B,C,D)$ for compactness, and noting that $\lambda_R' = \lambda_A' \lambda_C'$ we can rewrite Eq. \ref{Rank2} as
\begin{equation}
|\lambda_R' R| = |\lambda_R' R, +S, \lambda_R' T| + |\lambda_R' R, -S, -\lambda_R' T|.
\end{equation}
Rewriting the PPS as $|\Psi\rangle = |\lambda_E \lambda_G S\rangle$ and $|\Phi\rangle = |\lambda_I \lambda_K T\rangle$, we can see that the weak values will depend explicitly on the sign of $\lambda_S \equiv\lambda_E \lambda_G$ in the PPS.  If $\lambda_S = +1$, then only projectors with $\lambda_R'=\lambda_T \equiv \lambda_I \lambda_K$ will have nonzero weak values, while if $\lambda_S = -1$, then only projectors with $\lambda_R'= - \lambda_T$ will have nonzero weak values.  One can then see that the orbit around $|\mathcal{F}|$ is always nonzero because $\lambda_R' = \lambda_A' \lambda_C' = \lambda_S \lambda_T = \lambda_E \lambda_G\lambda_I \lambda_K$, while the orbit around $|\mathcal{G}|$ is always zero because $\lambda_R' = \lambda_A' \lambda_C' = -\lambda_S \lambda_T = - \lambda_E \lambda_G\lambda_I \lambda_K$.  The remaining orbit has the same four weak values as the 3-qubit Wheel, and the anomalous weak value always belongs to the projector of Eq. \ref{Anomaly}, which has all opposite eigenvalues to $|\mathcal{F}|$.\newline

This calculation can be straightforwardly generalized to the case of arbitrary even $N$ to obtain the formulae given above.

This example also reveals an important general restriction on conflict projectors for which the realist truth-value assignment $v_e$ is forced to be 0.  In a general set, it may be that the projectors in a conflict basis break up into several distinct orbits (each belonging to a KS subset), and each orbit can have its own magnitude of weak value.  If multiple orbits have nonzero weak value, then there is no guarantee that the weak values in each orbit will have a positive sum, and because of this, there is no general maximum weak value for a conflict projector in a basis with multiple nonzero orbits.

For a set and PPS with just one nonzero orbit in the conflict basis, we obtain the rule that the upper bound on the weak value of the conflict projectors is $w_{max} = 1/2$.  To see this, consider that there must be an even number $m$ of projectors in the orbit, all with the same magnitude weak value $w$, but with potentially arbitrary signs.  It is easy to see that the smallest possible positive sum of these $m$ weak values is $2w$, and the upper bound then follows from the sum rule.  Furthermore, it follows from the overall negative sign of a KS set, that every orbit must also contain at least one negative weak value.

The basis containing this orbit is always an eigenbasis of the conflict ID in the set.  There may be other hybrid conflict bases with all 0 realist truth-value assignments $v_e$ composed of projectors from several different eigenbases, as shown in Fig. \ref{Bases}.  As a result, these conflict bases may need not contain any of the negative weak values from conflict eigenbasis, but the upper bound of still applies.

\subsubsection{The 6-qubit Arch}  \label{ArchDerivation}

\begin{figure}[h!]
\centering
\caption{The 6-qubit Arch}
\includegraphics[width=3in]{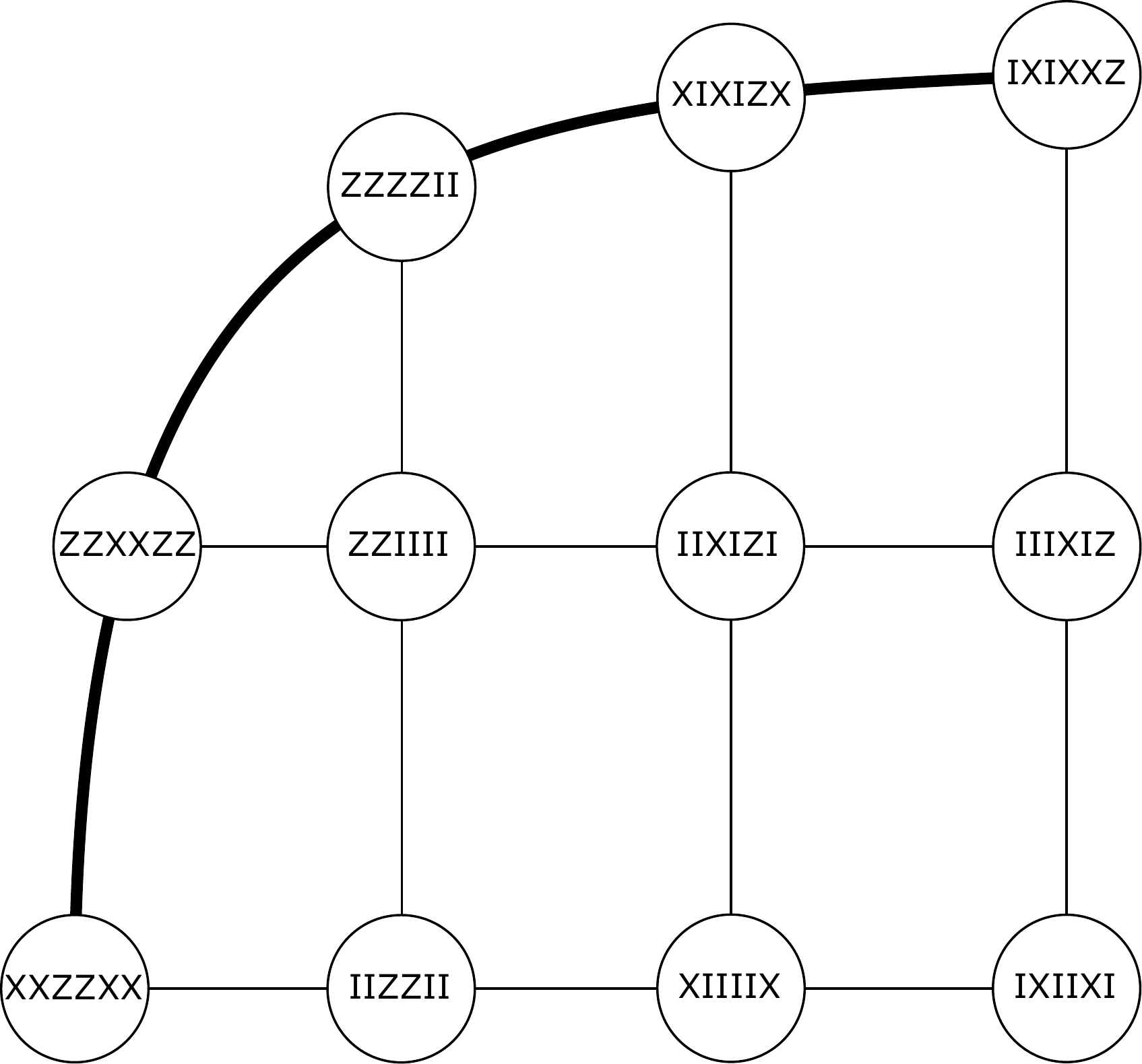}
\label{Arch}
\end{figure}

We work through one more KS set in order to make sure the general features are apparent.  The 6-qubit Arch of Fig. \ref{Arch} is a remarkably compact KS set that can also be used to show the quantum pigeonhole effect by choosing the PPS from among the two horizontal IDs and the curved ID.  We show the brief calculation for the weak values in the case that the conflict occurs in a product basis (which can be converted into the classical basis by local unitary operations).

To begin, we label the observables as in the other examples.  Starting from the bottom left, we label the curved ID in the order $(A,B,C,D,E)$.  The two horizontal IDs are then labeled from left to right in the order $(B,F,G,H)$ and $(A,I,J,K)$.

We choose as our pre-selection as $|\Psi\rangle = |\lambda_A A, \lambda_B B, \lambda_C C, \lambda_D D, \lambda_E E\rangle$, with $\lambda_A \lambda_B \lambda_C \lambda_D \lambda_E = s_{_{ABCDE}} = -1$, and our post - selection $|\Psi\rangle = | \lambda_B B, \lambda_F F, \lambda_G G, \lambda_H H\rangle$, with with $\lambda_B \lambda_F \lambda_G \lambda_H = s_{_{BFGH}} = +1$.

This forces the states $|f_1\rangle = |\lambda_C C,  \lambda_F F,  \lambda_C \lambda_F I \rangle$, $|f_2\rangle = |\lambda_D D,  \lambda_G G,  \lambda_D \lambda_G J \rangle$, and $|f_3\rangle = |\lambda_E E,  \lambda_H H,  \lambda_E \lambda_H K \rangle$.  These three forced states then induce a conflict basis, with the maximum conflict state given by,
\begin{equation}
|\mathcal{C}| = |\lambda_A A, -\lambda_C \lambda_F I, -\lambda_D \lambda_G J, -\lambda_E \lambda_H K|
\end{equation}
\begin{equation}
\equiv |\lambda_A A, \lambda_I I,  \lambda_J J,  \lambda_K K|,
\end{equation}
with $\lambda_A \lambda_I \lambda_J  \lambda_K = s_{_{AIJK}} = +1$ (which can be verified by direct substitution).  Notice also that in this case $\lambda_A$ is fixed by the PPS, and thus the weak value of projectors with $\lambda_A' = -\lambda_A$ is zero.  We now proceed to find the relation between the weak values of the other four projectors in the conflict basis,
\begin{equation}
|\lambda_A A, \lambda_I I,  \lambda_J J,  \lambda_K K|_w
\end{equation}
\begin{equation}
=\lambda_C \lambda_F \lambda_I s_{_{CFI}}|\lambda_A A, \lambda_I I,  -\lambda_J J,  -\lambda_K K|_w
\end{equation}
\begin{equation}
= \lambda_D \lambda_G \lambda_J s_{_{DGJ}}|\lambda_A A, -\lambda_I I,  \lambda_J J,  -\lambda_K K|_w
\end{equation}
\begin{equation}
=\lambda_E \lambda_H \lambda_K s_{_{EHK}} |\lambda_A A, -\lambda_I I,  -\lambda_J J,  \lambda_K K|_w,
\end{equation}
and note that $\lambda_C \lambda_F \lambda_I s_{_{CFI}} = \lambda_D \lambda_G \lambda_J s_{_{DGJ}} = \lambda_E \lambda_H \lambda_K s_{_{EHK}} = -1$.
So once again we find that the maximum conflict state $|\mathcal{C}|$ is the projector with the negative weak value, and in this case all four weak values have magnitude $1/2$.

Finally, the product of the signs of the all four weak values can be written as,
\begin{equation}
\lambda_C \lambda_D \lambda_E\lambda_F \lambda_G \lambda_H \lambda_I \lambda_J \lambda_K s_{_{CFI}}s_{_{DGJ}}s_{_{EHK}}
\end{equation}
\begin{equation}
=\lambda_A\lambda_B\lambda_C \lambda_D \lambda_E s_{_{CFI}}s_{_{DGJ}}s_{_{EHK}}s_{_{AIJK}}s_{_{BFGH}}
\end{equation}
\begin{equation}
=s_{_{CFI}}s_{_{DGJ}}s_{_{EHK}}s_{_{AIJK}}s_{_{BFGH}}s_{_{ABCDE}} = - 1,
\end{equation}
showing again that the negative weak value occurs exactly because this is a KS set.

\subsubsection{KS Sets and PPSs that do not lead to Conflict Bases}  \label{GHZStar}

As we have discussed, not every KS set of PPS leads to a conflict basis.

For example, consider the 3-qubit Square case we explored in detail.  If we had chosen a row and a column of the square as our PPS, instead of two different rows, there would be no forced values, and thus no conflict basis.

\begin{figure}[h!]
\caption{The 3-qubit GHZ-Mermin Star cannot give rise to a conflict basis.}
\centering
\subfloat[][No assignment]{
\includegraphics[width=1.25in]{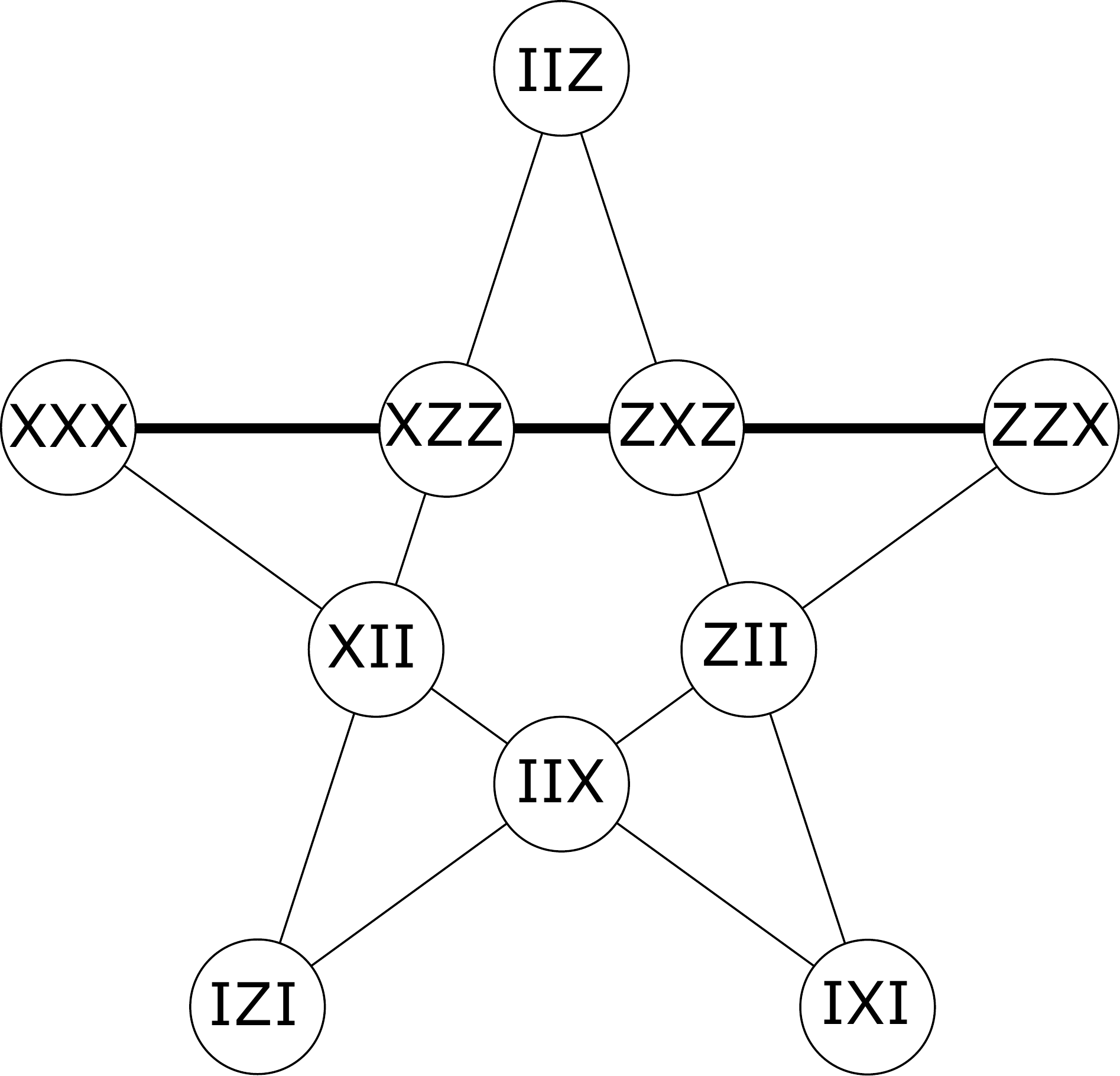}\label{Star}}
\qquad
\subfloat[][Pre- and post-selection]{
\includegraphics[width=1.25in]{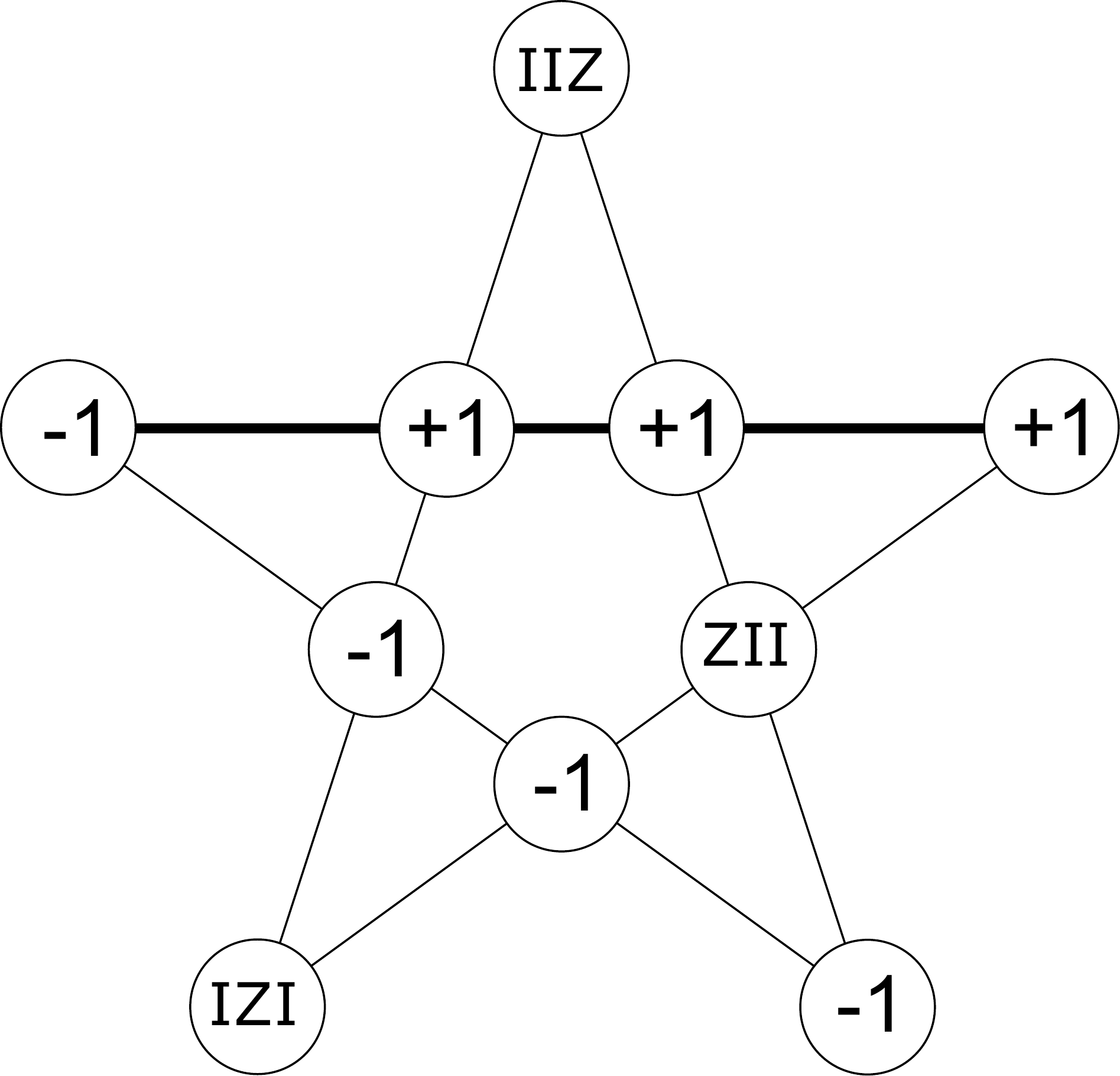}\label{Star1}}
\label{StarSteps}
\end{figure}

There are also KS sets that do not produce a conflict basis for any PPS, like the 3-qubit GHZ-Mermin Star of Fig. \ref{Star}.  It is easy to see that the PPS shown in Fig. \ref{Star1} is symmetric to any other choice of PPS from within this set, and does not force any additional values.

\begin{figure}[h!]
\caption{(Color Online) This modified version of the GHZ-Mermin star does give rise to a conflict basis for the same PPS as in Fig. \ref{StarSteps}.}
\centering
\subfloat[][No assignment]{
\includegraphics[width=1.25in]{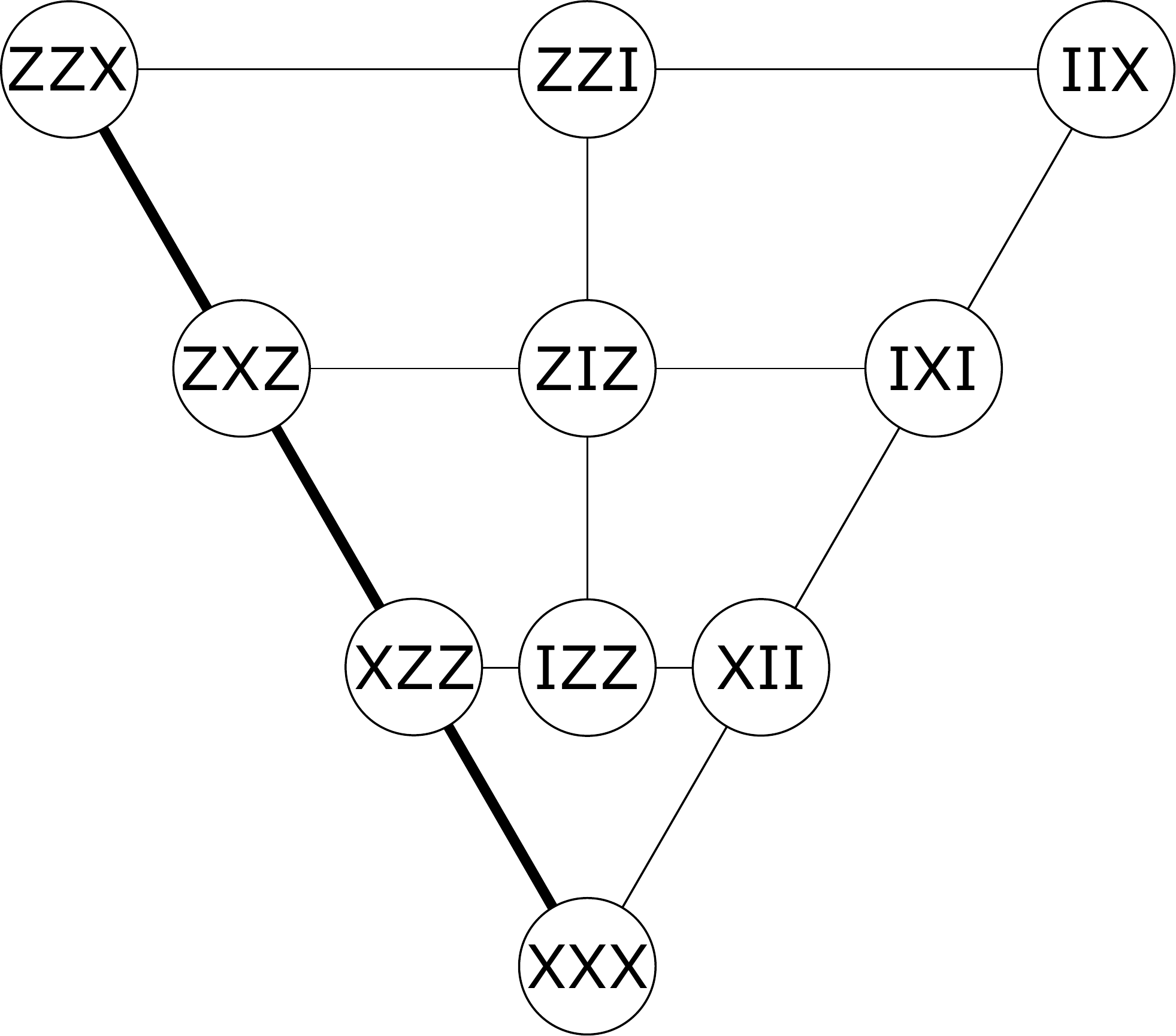}\label{StarPigeon}}
\qquad
\subfloat[][Pre- and post-selection]{
\includegraphics[width=1.25in]{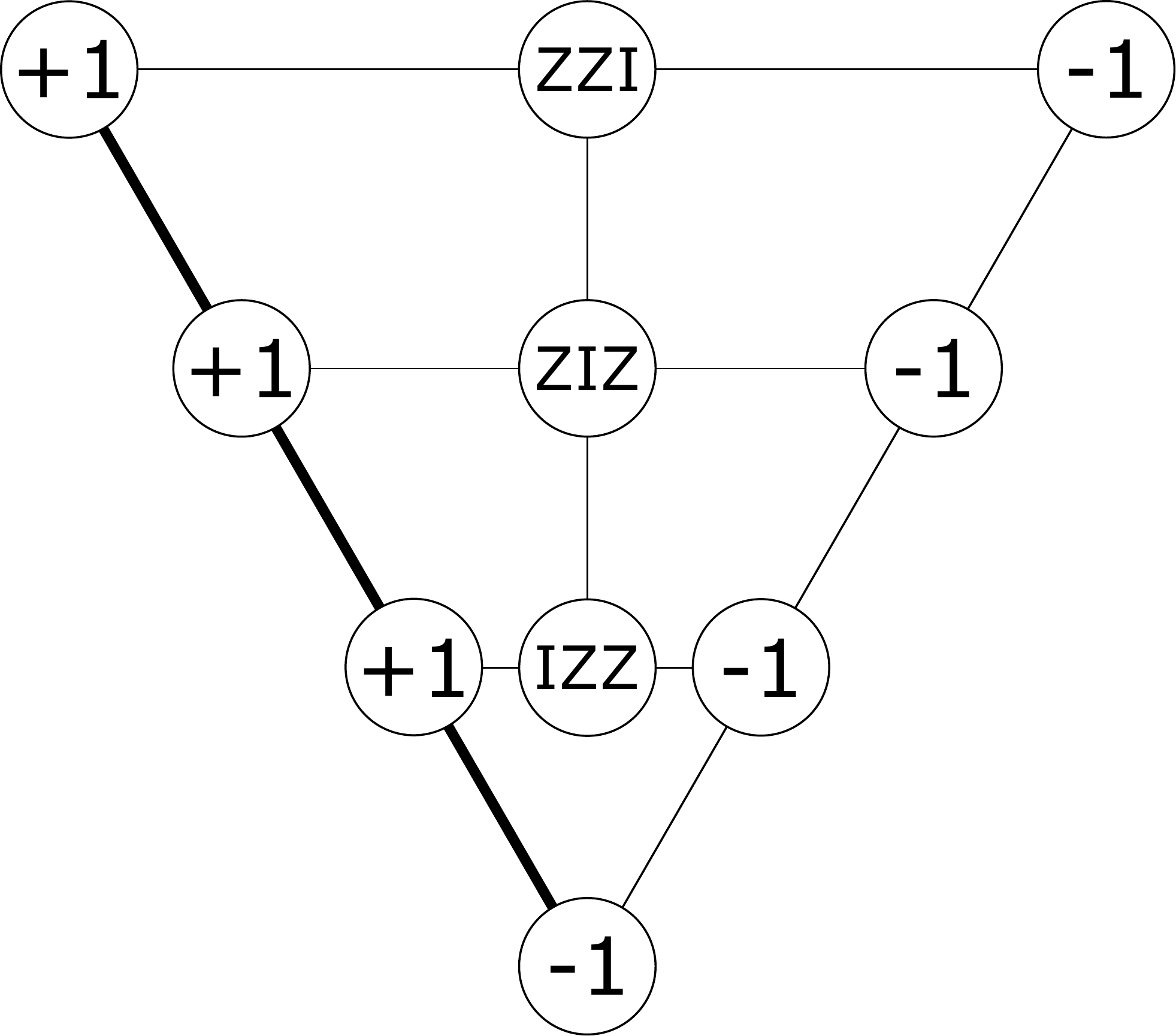}\label{StarPigeon1}}
\qquad
\subfloat[][The ABL rule and violation of the product rule (shown by the blue dashed line)]{
\includegraphics[width=1.25in]{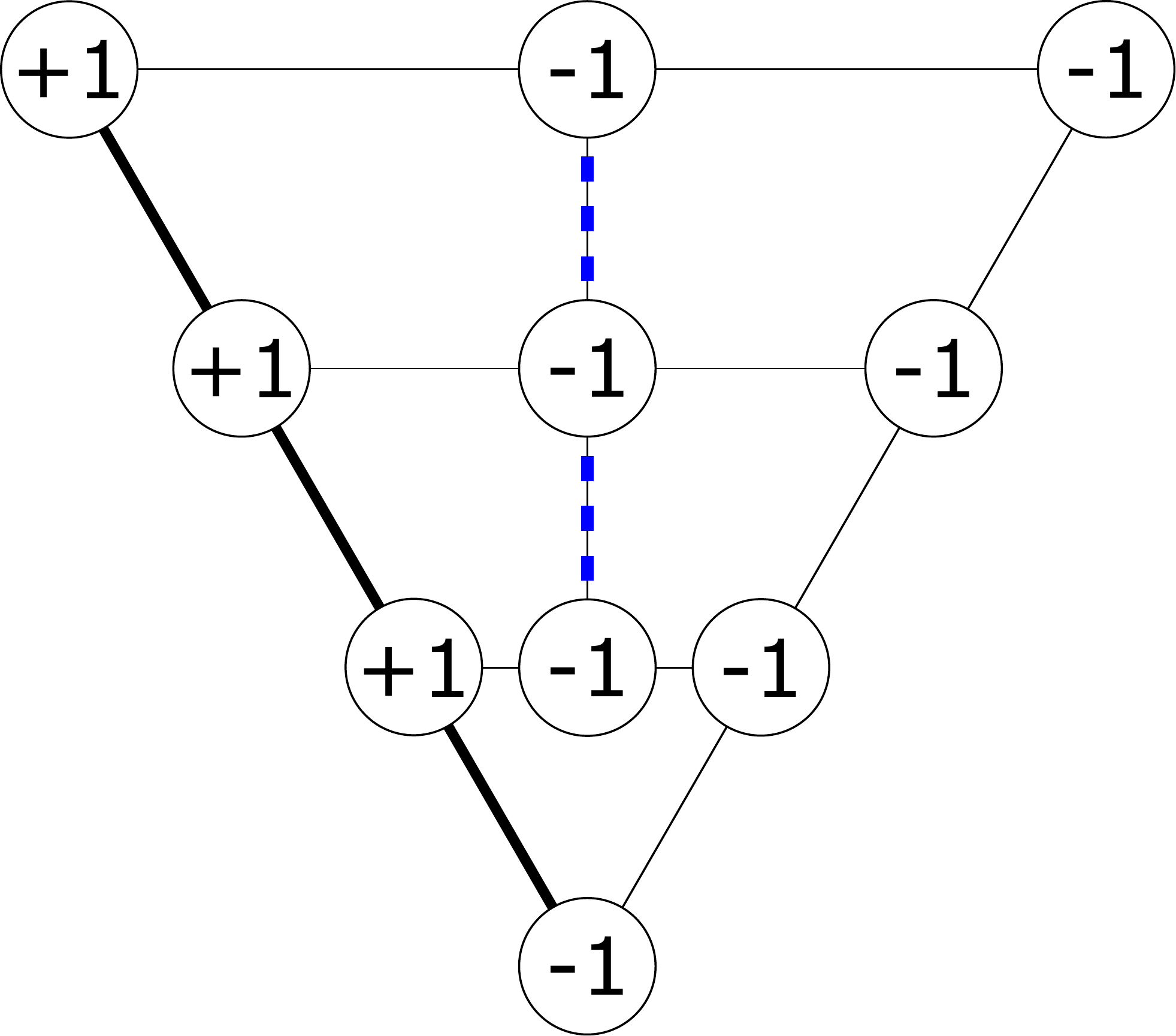}\label{StarPigeon2}}
\label{StarPigeonSteps}
\end{figure}

This is not to say that this PPS does not give rise to any conflict bases in the entire 3-qubit Pauli group, only that those contexts do not belong to this KS set.  With a fairly simple modification of this KS set, we can obtain a different KS set that does produce a conflict basis for this same PPS.  The modified set and the steps leading to the conflict are shown in Fig. \ref{StarPigeonSteps}.  Again, we find the conflict in the classical basis, and we have the pigeonhole effect, but in this case our PPS included a maximally entangled GHZ state, so this is not the product pigeonhole effect.

This conflict basis was first obtained using this PPS by Tollaksen \cite{tollaksen2007pre}.

\subsubsection{The $N$-qubit Quantum Pigeonhole Effect Without a KS Set} \label{NonKS}

Throughout this paper we have focussed on the connection between the quantum pigeonhole effect and KS sets, but the KS theorem is not the only demonstration of quantum contextuality, and likewise it is not the only demonstration of the quantum pigeonhole principle.

In this section we give an $N+1$-qubit ($N$ even) proof of quantum contextuality that uses product states for the PPS, and forces a conflict in the classical basis (i.e. the same trick we can play with the $N$-qubit Wheels), but does not depend on any complete KS set that we can identify.  We were made aware of this proof by Yakir Aharonov, and what follows is truly only a reproduction and slight generalization of his ideas.

For this proof, we consider the positive ID of Fig. \ref{IDPigeon}, which has the classical basis as its eigenbasis.
 \begin{figure}[h!]
 \centering
 \caption{An $(N+1)$-qubit positive ID with $N+1$ observables}
 \subfloat[][$N=4$]{
 \begin{tabular}{ccccc}
 $I$ & $Z$ & $Z$ & $Z$ & $Z$ \\
 $Z$ & $I$ & $Z$ & $Z$ & $Z$ \\
 $Z$ & $Z$ &   $I$ & $Z$ &  $Z$ \\
$Z$ & $Z$ & $Z$ & $I$ & $Z$ \\
 $Z$ & $Z$ & $Z$ & $Z$ & $I$ \\
 \end{tabular}}
 \qquad
 \subfloat[][Any even $N$]{
 \begin{tabular}{ccccc}
 $I$ & $Z$ & $Z$ & $\cdots$ & $Z$ \\
 $Z$ & $I$ & $Z$ & $\cdots$ & $Z$ \\
 $\vdots$ &  $\ddots$ &   $\ddots$ & $\ddots$ &  $\vdots$ \\
$Z$ & $\cdots$ & $Z$ & $I$ & $Z$ \\
 $Z$ & $\cdots$ & $Z$ & $Z$ & $I$ \\
 \end{tabular}\label{IDPigeon}}
 \end{figure}

 We will choose product states as our pre- and post-selected states, and thus we can consider each qubit individually, and then multiply the results together.  To begin we consider the case that each qubit is pre-selected in the state $|\Psi\rangle = |+X\rangle$, and post-selected in the state $|\Phi\rangle = \cos{(n\pi/2N)}|0\rangle - i \sin{(n\pi/2N)}|1\rangle$, where $n$ is any odd integer and $\{|0\rangle \equiv |+Z\rangle,|1\rangle \equiv |-Z\rangle\}$ is the computational basis.  The weak value of each single-qubit Pauli $Z$ observable is then $Z_w = e^{in\pi /N}$.  It then follows that the weak value of each $(N+1)$-qubit observable in Fig.\ref{IDPigeon} is -1.  Using the reverse ABL rule to force these values, we find a conflict in the classical basis, since Fig.\ref{IDPigeon} is a positive ID.

  We can also show that these cases always give rise to projectors with negative real parts for their weak values, which is another direct verification of contextuality.  Specifically we consider the maximum conflict projector, $|\mathcal{C}| \equiv |\mathcal{C}\rangle\langle\mathcal{C}|$, that maximally opposes the PPS by having eigenvalue +1 for all $N+1$ observables.  This is actually a rank-2 projector, and therefore we first decompose it into a sum of orthogonal rank-1 projector as follows,
 \begin{equation}
 |\mathcal{C}| = |+Z_1, +Z_2, \ldots, +Z_{N+1}| + |-Z_1, -Z_2, \ldots, -Z_{N+1}|.
 \end{equation}
 Each term in this sum can be factored into the tensor product of $|\pm Z|$ for each individual qubit, and thus the weak value of the projector is,
 \begin{equation}
 |\mathcal{C}|_w = e^{ -i n\pi (N+1) /2N} [(\cos{\frac{n\pi}{2N}})^{N+1} + ( i \sin{\frac{n\pi}{2N}})^{N+1}],
 \end{equation}
 which has a negative real part that approaches zero as $N\rightarrow\infty$.  It is also noteworthy that the weak values of the classical projectors generally do {\it not} have the same magnitudes with this PPS, as they do with KS sets.  Thus, the uniform ABL probabilities for the projectors in a conflict basis appears to be a feature that is unique to KS sets - perhaps due to their additional symmetry properties.

 To extend this proof to full generality, we will now consider any pre-selected state in the XZ plane, $|\Psi\rangle = \cos{(\theta/2)}|0\rangle + \sin{(\theta/2)}|1\rangle$, which combined with the (unnormalized) post-selected state $|\Phi\rangle = \sin{\frac{\theta}{2}}\cos{\frac{n\pi}{2N}}|0\rangle - i \cos{\frac{\theta}{2}}\sin{\frac{n\pi}{2N}}|1\rangle$ (which is in the YZ plane), we again obtain $Z_w = e^{in\pi /N}$, which leads to the conflict just as before.

 The proof we have shown here fails for odd $N$, but just as with the Wheels, the proof can be recovered by considering all of the different $(N-1)$-qubit IDs like Fig.\ref{IDPigeon} that fall within the classical stabilizer group.

 As a final generalization, note that each observable of Fig.\ref{IDPigeon} is composed of $N$ Pauli $Z$ observables and one single-qubit identity ($I$) operator.  There exists a much more general set of IDs from within the classical stabilizer group for which the above PPSs will work, and these are IDs composed of $N$ Pauli-$Z$ observables for any even $N$, and $m$ single-qubit identity ($I$) operators, for any odd $m$.  From symmetry, these positive IDs must contain an odd number of observables, and thus the PPS (for all $N+m$ qubits) forces a conflict in the classical basis.

\end{document}